\documentclass[aps, prd, 10pt, twocolumn, superscriptaddress, noshowpacs, preprintnumbers, longbibliography, groupedaddress, footinbib, bibnotes]{revtex4-1}

\usepackage{amsmath}
\usepackage{amsfonts}
\usepackage{amssymb}
\usepackage{bbold}
\usepackage{epsfig}
\usepackage{graphicx}
\usepackage{bm}
\usepackage{array}
\usepackage{hyperref}
\usepackage{listings}
\usepackage{color}
\usepackage[normalem]{ulem}

\newcommand{\Hvac}{H_{\textrm{vac}}}
\newcommand{\Hmat}{H_{\textrm{mat}}}
\newcommand{\Hnunu}{H_{\nu\nu}}
\newcommand{\Closs}{\mathcal{C}_{\textrm{loss}}}
\newcommand{\Cgain}{\mathcal{C}_{\textrm{gain}}}
\newcommand{\Clossbar}{\bar{\mathcal{C}}_{\textrm{loss}}}
\newcommand{\Cgainbar}{\bar{\mathcal{C}}_{\textrm{gain}}}

\begin{document}

\title{A change of direction in pairwise neutrino conversion physics: The effect of collisions}

\author{Shashank Shalgar}
\email{shashank.shalgar@nbi.ku.dk}
\thanks{ORCID: \href{http://orcid.org/0000-0002-2937-6525}{0000-0002-2937-6525}}
\affiliation{Niels Bohr International Academy \& DARK, Niels Bohr Institute,\\University of Copenhagen, 2100 Copenhagen, Denmark}

\author{Irene Tamborra}
\email{tamborra@nbi.ku.dk}
\thanks{ORCID: \href{http://orcid.org/0000-0001-7449-104X}{0000-0001-7449-104X}}
\affiliation{Niels Bohr International Academy \& DARK, Niels Bohr Institute,\\University of Copenhagen, 2100 Copenhagen, Denmark}

\date{\today}

\begin{abstract}
Fast pairwise conversions of neutrinos may affect the flavor distribution in the core of neutrino-dense sources. We explore the interplay between collisions and fast conversions within a simplified framework that assumes angle-independent, direction-changing collisions in a neutrino gas that has no spatial inhomogeneity. Counter to expectations, we find that collisions may enhance fast flavor conversions instead  of damping them. Our work highlights the need to take into account the feedback of collisions on the neutrino angular distributions self-consistently, in order to predict the flavor outcome in the context of fast pairwise conversions reliably.

\end{abstract}

\maketitle

\section{Introduction}
\label{sec:intro}

Neutrino flavor evolution in dense media, such as core-collapse supernovae, compact binary mergers, and the early universe, is one of the most fascinating macroscopic quantum mechanical phenomena 
far from being understood~\cite{Mirizzi:2015eza,Duan:2010bg,Chakraborty:2016yeg}. In particular, one of the least explored topics in the context of neutrino mixing concerns the interplay between neutrino flavor conversions and scatterings.

The coherent forward scattering of neutrinos with the matter background can lead to a resonant enhancement of neutrino flavor conversions as first pointed out by Mikheev, Smirnov, and Wolfenstein~\cite{Mikheev:1986if,1985YaFiz..42.1441M,1978PhRvD..17.2369W}. Similarly, the flavor conversion history can be affected by the non-linear feedback induced by the forward scattering of neutrinos among themselves~\cite{1987ApJ...322..795F,Notzold:1987ik,Pantaleone:1992eq,Pantaleone:1992xh,Duan:2010bg,Mirizzi:2015eza}. 

The possibility that the flavor evolution history may be affected by residual collisions with the background medium was discussed in Refs.~\cite{Stodolsky:1974hm,Harris:1980zi,Stodolsky:1986dx}. It was pointed out that the random scattering of neutrinos on a stochastic background medium, in the case of a high collision strength, could destroy coherence in the flavor evolution; in this case, the impact of collisions is dictated by the damping rate that measures the degree at which the coherent development of the neutrino wavefunction is halted by collisions. 
The final flavor outcome is then connected to the relative ratio between the effective neutrino conversion rate and the collisional rate. 

The role of collisions has been widely investigated in the early universe, where it has been found that collisions play a vital role in the development of flavor asymmetries that, in turn, are essential to the flavor evolution. Collisions are taken into account in the early universe by using the Boltzmann statistics for neutrinos~\cite{Hannestad:1995rs,Dolgov:1997mb,Dolgov:1998sf,Esposito:2000hi}, along with a damping approximation for the off-diagonal terms of the density matrices~\cite{deSalas:2016ztq}. Collisions may lead to spectral distortions of the neutrino energy distributions which, in turn, may affect the relic density of neutrinos~\cite{Mangano:2005cc}.

In the context of core-collapse supernovae, the neutrino mixing between active flavors has been considered to happen at relatively low densities, when neutrinos are in the free streaming regime, and collisions were deemed to play a negligible role in the context of $\nu$--$\nu$ interactions~\cite{Sigl:1992fn,Duan:2005cp,Duan:2006jv,Vlasenko:2013fja}. However, Ref.~\cite{Cherry:2012zw} pointed out that, despite the fact that only a small fraction of neutrinos undergoes direction-changing scattering in the free streaming regime, this may still impact the overall flavor outcome; neutrinos occasionally scattering on the matter envelope could broaden their angular distribution with respect to the one usually assumed by considering neutrinos streaming out of the supernova core (the so-called ``neutrino halo''). The neutrino halo was found to dominate $\nu$--$\nu$ refraction at distances larger than $\mathcal{O}(100)$~km from the core~\cite{Cherry:2012zw}. By employing the linear stability analysis, Ref.~\cite{Sarikas:2012vb} concluded that multi-angle matter suppression~\cite{EstebanPretel:2008ni} of neutrino self-interactions still holds in the presence of the neutrino halo. Several other attempts of understanding the role of the neutrino halo in the context of flavor conversions involve various degrees of approximation~\cite{Cherry:2013mv,Cirigliano:2018rst,Zaizen:2019ufj,Cherry:2019vkv}; hence, to date, a self-consistent assessment of the role of the neutrino halo in the flavor mixing is lacking.

More recently, it has been pointed out that flavor mixing can also occur at much higher densities because of pairwise neutrino scattering, leading to fast  conversions~\cite{Sawyer:2015dsa,Sawyer:2005jk,Sawyer:2008zs,Chakraborty:2016yeg,Chakraborty:2016lct,Izaguirre:2016gsx}. One of the conditions identified as responsible for triggering fast pairwise conversions is the presence of an effective crossing between the angular distributions of $\nu_e$ and $\bar\nu_e$ (ELN crossing)~\cite{Izaguirre:2016gsx,Yi:2019hrp,Martin:2019gxb,Abbar:2017pkh}. 

The sensitivity of pairwise conversions to the exact shape of the angular distributions suggests that collisions may play a fundamental role in preparing the neutrino gas before fast pairwise conversions. In fact, in the supernova core, the matter density is so large that neutrinos are trapped and their angular distribution is isotropic. As the radius increases and the matter density falls, first the angular distribution of non-electron type neutrinos starts becoming forward peaked, followed by the one of electron 
antineutrinos and  neutrinos~\cite{Brandt:2010xa,Tamborra:2017ubu}. By exploring the conditions under which ELN crossings form because of collisions, it was found that they tend to occur when the number density of $\nu_e$ approaches the one of $\bar\nu_e$: $n_{\nu_e}/n_{\bar{\nu}_e} \simeq 1$~\cite{Shalgar:2019kzy}. Hence, as the angular distribution of neutrinos starts becoming forward peaked, ELN crossings arise~\cite{Tamborra:2017ubu,Shalgar:2019kzy,Abbar:2019zoq, Abbar:2018shq,DelfanAzari:2019tez,Nagakura:2019sig,Morinaga:2019wsv,Wu:2017drk,Wu:2017qpc,Glas:2019ijo,Xiong:2020ntn,George:2020veu,Padilla-Gay:2020uxa}.

The rate of fast neutrino conversions is proportional to $\mu = \sqrt{2} G_F n_\nu$, with $G_F$ being the Fermi constant and $n_{\nu}$ the total (anti)neutrino density. 
Throughout this paper, we use $\mu=10^{5}$ km$^{-1}$, hence the growth rate of the flavor instability associated to fast flavor conversions is $\mathcal{O}(10^{2}$--$10^{3})$ km$^{-1}$, while the collision strength is $\mathcal{O}(0.1$--$1)$~km$^{-1}$. Given the clear hierarchy among characteristic scales, naively one would not expect collisions to have any impact on flavor evolution, but as we will see our results contradict such expectations.
The potential relevance of collisions in this context was discussed in Ref.~\cite{Capozzi:2018clo} where, by employing a one-dimensional model with two momentum modes, it was shown that collisions create favorable conditions for flavor conversions to be triggered.

In this paper, we focus  on the high density region, e.g.~correspondent to the core of supernovae or compact binary mergers, and by employing a simple benchmark model explore the role of collisions on the flavor mixing. Given the major challenges induced by a self-consistent implementation of the flavor mixing in the presence of collisions, we adopt a very simple framework with the direction-changing collisional term having the same intensity for all neutrino momenta.  We assume three different angular distributions and then explore how they are modified by the interplay between collisions and flavor mixing. However, we do not let collisions shape the flavor distributions, given the simplified treatment of our collisional term. Our goal is not to make self-consistent predictions of the flavor mixing in the presence of collisions, but to explore the new effects in the flavor mixing phenomenology determined by collisions.

This paper is organized as follows. In Sec.~\ref{sec2}, we introduce the physics of fast neutrino conversions and motivate the setup adopted for our  initial conditions. In Sec.~\ref{sec3} we present our simplified implementation of the scatterings of neutrinos with nucleons. The interplay between collisions and flavor conversions is explored in Sec.~\ref{sec:results} when the collisional term for neutrinos is identical to the one of antineutrinos and when they differ from each other. In Sec.~\ref{sec:conc}, we discuss and summarize our findings.

\section{Neutrino equations of motion}
\label{sec2}

We can represent the neutrino field in terms of Wigner transformed density matrices, which we denote by $\rho$ (or $\bar{\rho}$ for antineutrinos). 
The flavor evolution can be schematically expressed as follows:
\begin{eqnarray}
\label{eom1}
\left(\frac{\partial}{\partial t} + \vec{v}\cdot\nabla\right) \rho(\vec{p}) =
-i [H(\vec{p}),\rho(\vec{p})] + \mathcal{C}[\rho,\bar{\rho}]\ ,\\
\left(\frac{\partial}{\partial t} + \vec{v}\cdot\nabla\right) \bar{\rho}(\vec{p}) =
-i [\bar{H}(\vec{p}),\bar{\rho}(\vec{p})] + \bar{\mathcal{C}}[\rho,\bar{\rho}]\ .
\label{eom2}
\end{eqnarray}
The Hamiltonian includes the vacuum term and the contribution due to the coherent forward scattering, while $\mathcal{C}$ and $\bar{\mathcal{C}}$ represent the collision term due to the incoherent part of the scattering, which is discussed in Sec.~\ref{sec3}. The left-hand side of Eqs.~\ref{eom1} and \ref{eom2} is the covariant derivative which includes the explicit time evolution as well as the advective term. Moreover, we neglect neutrino advection in Eqs.~\ref{eom1} and \ref{eom2}; this implies that we are considering a system in which the spatial derivatives vanish, although it has been shown that neutrino advection may smear the ELN crossings hindering the development of flavor conversions~\cite{Shalgar:2019qwg}. 

In most studies on neutrino mixing in dense media, simplifications have to be made in order to make the problem computationally feasible, see Refs.~\cite{Duan:2010bg,Mirizzi:2015eza} for reviews on the topic. 
For ultrarelativistic neutrinos  with momentum $\vec{p}$, the Hamiltonian consists of three terms
\begin{eqnarray}
H(\vec{p}) &=& \Hvac + \Hmat + \Hnunu \\
\bar{H}(\vec{p}) &=& -\Hvac + \Hmat + \Hnunu\ , 
\end{eqnarray}
the vacuum term (which has opposite sign for neutrinos and antineutrinos) and the contribution from coherent forward scattering terms with the matter and neutrino backgrounds. The three terms of the Hamiltonian are defined as follows
\begin{eqnarray}
\Hvac &=& 
\frac{\omega}{2} \begin{pmatrix} 
-\cos 2 \theta_{\textrm{V}} & \sin 2 \theta_{\textrm{V}}\\
\sin 2 \theta_{\textrm{V}} & \cos 2 \theta_{\textrm{V}}
\end{pmatrix}\\
\Hmat &=&
\begin{pmatrix} 
\sqrt{2}G_{\textrm{F}}n_{e} & 0\\
0 & 0
\end{pmatrix}\\
\Hnunu &=& 
\mu \int d^{3}p^{\prime} \left[\rho(\vec{p^{\prime}}) - \bar{\rho}(\vec{p^{\prime}})\right]
\left(1-\vec{v}\cdot \vec{v^{\prime}}\right),
\end{eqnarray}
where, $\omega = \frac{\Delta m^{2}}{2E}$ is the vacuum frequency, with $\Delta m^{2}$ corresponding to the neutrino mass difference and $E$ the neutrino energy, $\theta_{\textrm{V}}$ is the vacuum mixing angle, $n_{e}$ is the electron number density, $\mu = \sqrt{2} G_{\textrm{F}} n_{\nu_{e}}$ is the self-interaction potential, and $\vec{v}=\vec{p}/|\vec{p}|$. 

In what follows, for simplicity, we ignore the matter term in the Hamiltonian, $\Hmat$, as it only plays the role of reducing the magnitude of the effective mixing angle; however, we set $\theta_{\textrm{V}} = 10^{-6}$ to take into account the effective mixing suppression due to matter~\cite{EstebanPretel:2008ni}. Throughout the paper, we assume 
$\mu = 10^5$~km$^{-1}$.
 
The neutrino flavor conversion phenomenology due to $\nu$--$\nu$ interactions can be broadly divided into two types: slow and fast neutrino conversions. Slow conversions are expected to occur at smaller neutrino densities than fast conversions and depend on the interplay between $\omega$ and $\mu$~\cite{Duan:2010bg,Mirizzi:2015eza}. Fast pairwise conversions should instead occur at very large neutrino densities and are driven by $\mu$; in principle, fast conversions occur for $\omega =0$ in the presence of other perturbations that can trigger flavor mixing, although $\omega$ is known to affect the development of flavor conversions in the non-linear regime~\cite{Izaguirre:2016gsx,Johns:2019izj,Airen:2018nvp,Chakraborty:2016yeg,Dasgupta:2017oko,Johns:2020qsk,Shalgar:2020xns}.

\begin{figure}[t!]
\includegraphics[width=0.49\textwidth]{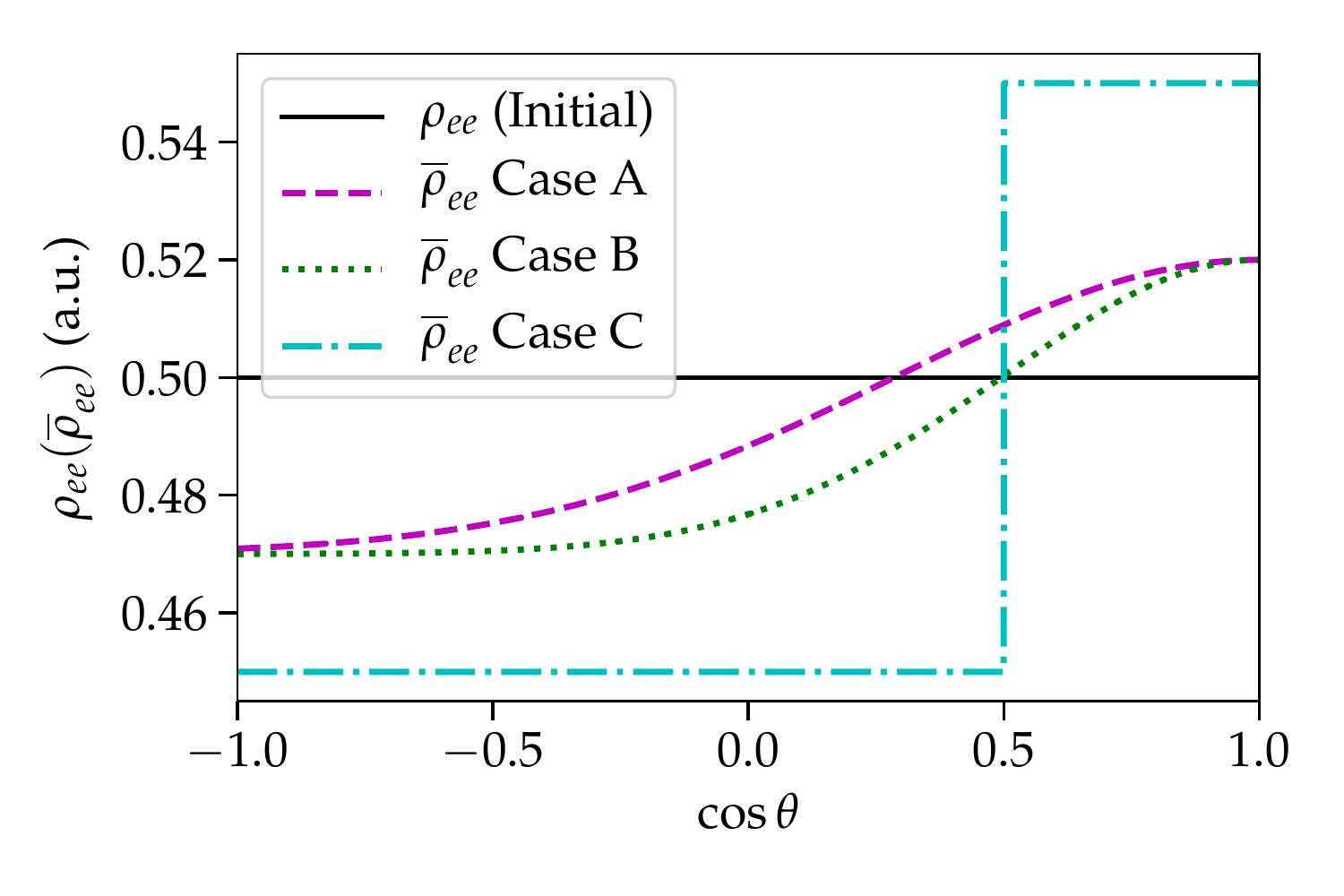}
\caption{Angular distributions of $\nu_e$ and $\bar\nu_e$ as functions of $\cos\theta$ adopted as inputs for our simulations. Three different configurations are explored (Cases A, B, and C); the $\nu_e$ angular distribution for neutrinos is the same for all three Cases and is shown by the black solid line, the angular distributions of $\bar\nu_e$ are different for each Case and are defined in Eq.~\ref{cases1}--\ref{cases3}.}
\label{Fig1}
\end{figure}

In what follows, we also assume axial symmetry and track the flavor evolution in the two flavor basis spanned by $(\nu_e, \nu_x)$ with $x=\mu, \tau$. 
We consider three illustrative sets of angular distributions for neutrinos and antineutrinos motivated by the conditions occurring near the decoupling region: Cases A, B, and C. All three Cases have an isotropic distribution for $\nu_e$'s, and different shapes for $\bar\nu_e$: 
\begin{subequations}
\begin{eqnarray}
\label{cases1}
& & \rho_{ee,\textrm{Case A}}(\cos\theta) = 0.5 \nonumber \\ 
 & & \bar{\rho}_{ee, \textrm{Case A}} = 0.47 + 0.05 \exp\left(-(\cos\theta-1)^{2}\right) \\
 \label{cases2}
 & & \rho_{ee, \textrm{Case B}}(\cos\theta) = 0.5 \nonumber \\ 
 & & \bar{\rho}_{ee, \textrm{Case B}} = 0.47 + 0.05 \exp\left(-2(\cos\theta-1)^{2}\right) \\
 & & \rho_{ee,\textrm{Case C}}(\cos\theta) = 0.5 \nonumber \\
 & & \bar{\rho}_{ee,\textrm{Case C}} = 
 \begin{cases} 0.45 \ & \textrm{ for } \cos\theta<0.5\ \\ 
 0.55\ & \textrm{ otherwise\ ;} 
 \end{cases}
\label{cases3}
\end{eqnarray}
\end{subequations}
the non-electron neutrinos are  generated through flavor mixing only. Cases A, B, and C are represented in Fig.~\ref{Fig1}. 

\section{Collision term}
\label{sec3}

Collisions are taken into account in Eqs.~\ref{eom1} and \ref{eom2} through the terms depending on $\mathcal{C}$ and $\bar{\mathcal{C}}$.
If collisions of neutrinos on nucleons are flavor blind, their main effect is to modify the momentum of neutrinos (both in direction and magnitude). When collisions are flavor dependent, they also have the effect of modifying the flavor ratio. In this work, for the sake of simplicity, we focus on 
collisions that are flavor blind and assume that collisions can only change the neutrino direction and not the energy. This is a valid approximation in the limit of small neutrino energy compared to the nucleon mass.

Our collision term is thus solely determined by the probability of a neutrino traveling in one direction and then changing direction, while maintaining the phase information. For each given angle bin, there is a finite probability that neutrinos will be scattered out of that angle bin, and a finite probability that neutrinos traveling in another direction will be scattered into that angle bin. These two effects are encoded in $\mathcal{C}$ and $\bar{\mathcal{C}}$ through the ``loss'' and ``gain'' terms. We also further simplify the system, by assuming mono-energetic neutrinos. This means that Eqs.~\ref{eom1} and \ref{eom2} can be written as
\begin{eqnarray}
\frac{d \rho(\cos\theta)}{dt} = &-& \int_{-1}^{1} \Closs \rho(\cos\theta) d\cos\theta^{\prime}\nonumber\\
&+& \int_{-1}^{1} \Cgain \rho(\cos\theta^{\prime}) d\cos\theta^{\prime}\nonumber\\
&-& i[H(\cos\theta),\rho(\cos\theta)]\ ,\nonumber\\
\frac{d \bar{\rho}(\cos\theta)}{dt} = &-& \int_{-1}^{1}\Clossbar \bar{\rho}(\cos\theta) d\cos\theta^{\prime}\nonumber\\
&+& \int_{-1}^{1} \Cgainbar \bar{\rho}(\cos\theta^{\prime}) d\cos\theta^{\prime}\nonumber\\
&-& i[\bar{H}(\cos\theta),\rho(\cos\theta)]\ .
\label{eoms}
\end{eqnarray}
 The characteristic length scales associated to collisions are $\Closs(\Clossbar)$ and $\Cgain(\Cgainbar)$, which are equal in this paper for simplicity; thus, we assume that the number of neutrinos and antineutrinos is conserved and $\int d\rho/dt d\cos\theta = \int d\bar\rho/dt d\cos\theta = 0$. 
 The term $\int \Closs (\Clossbar) d\cos\theta^{\prime} = 2\Closs(2\Clossbar)$ denotes the inverse of the mean-free-path.  In what follows, we indicate $\mathcal{C} = 2\Closs = 2\Cgain$ and $\bar{\mathcal{C}} = 2\Clossbar = 2\Cgainbar$ for neutrinos and antineutrinos, respectively. In addition, although we impose the conservation of total number density, Eqs.~\ref{eoms} are not unitary for any given angle, if $\mathcal{C} \not = 0$.

It should be noted that the first term on the right-hand-side of Eq.~\ref{eoms} is an integration that can be performed trivially as the integrand is independent of $\cos\theta^{\prime}$, but we write it in this form for two reasons: First, it lets us write the equations of motions such that the equality between $\Closs$ and $\Cgain$ implies particle number density conservation. Second, explicitly performing this trivial integration numerically in the  simulations aids  the numerical stability of the code by introducing the same magnitude or number error in the loss and the gain term.

\section{Impact of collisions on fast pairwise conversions}
\label{sec:results}
The numerical evolution of the neutrino flavor field is challenging in the case of fast pairwise conversions, as in most cases the angular distribution of neutrinos acquires finer structures as time increases. 
For all our results, we use $2000$ angle bins, which we have verified that are sufficient to achieve numerical convergence for the temporal range presented in this work. 
In the following, we first discuss the interplay between collisions and flavor conversions for the same collision strength between neutrinos and antineutrinos and then explore the case when the neutrino collision strength is different  from the one of antineutrinos.

\subsection{Identical collision strength for neutrinos and antineutrinos}

\begin{figure*}[t!]
\includegraphics[width=0.49\textwidth]{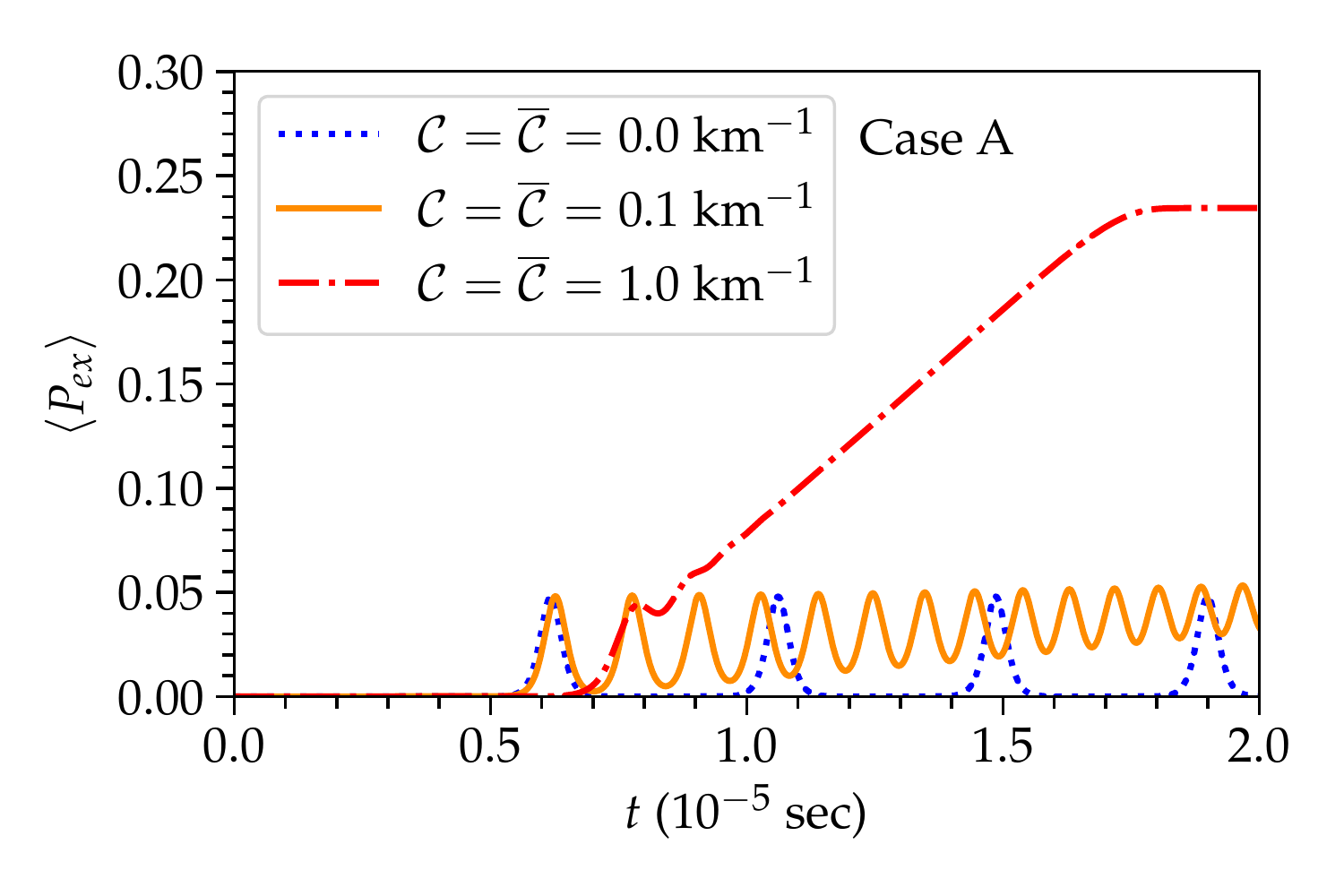}
\includegraphics[width=0.49\textwidth]{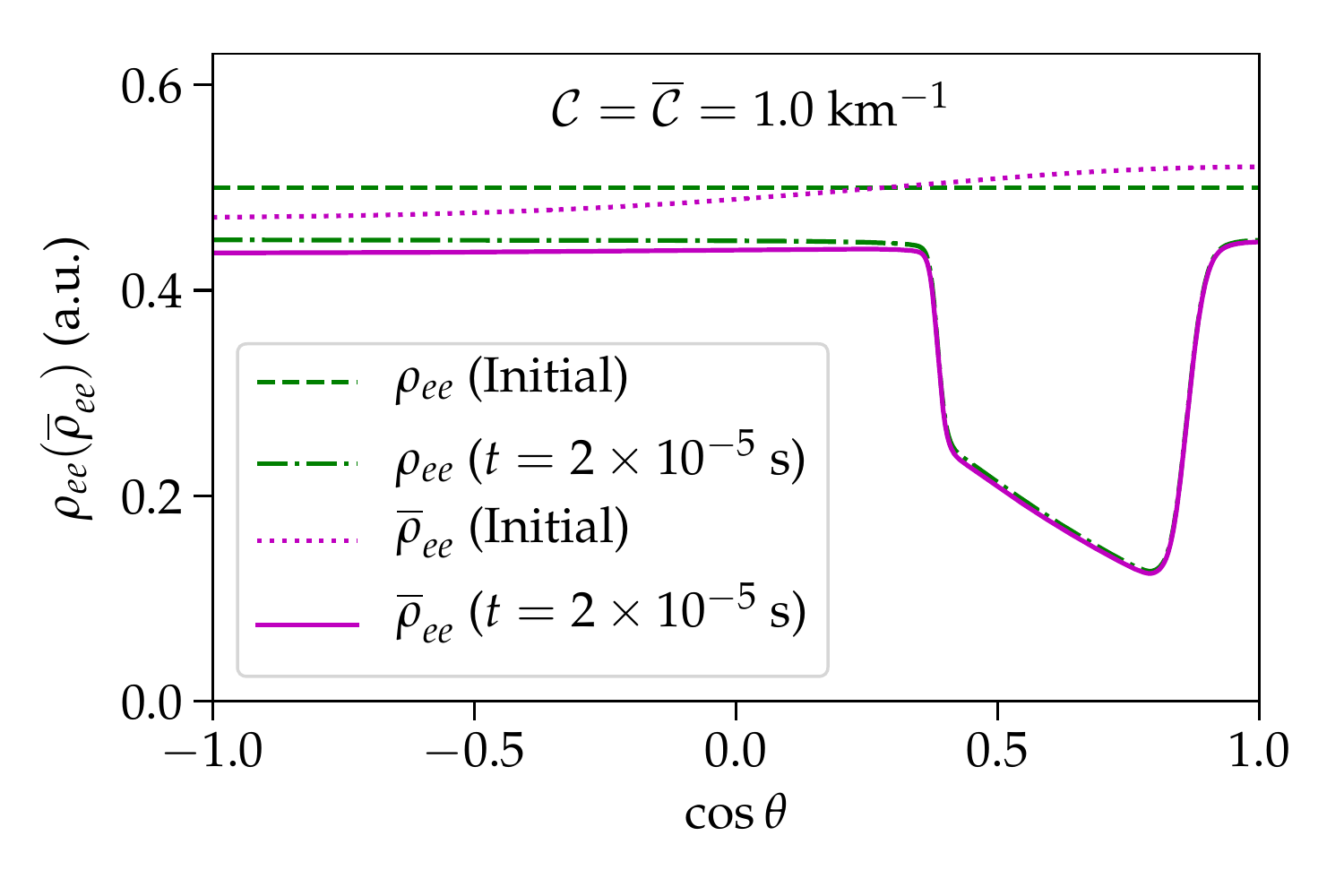}
\includegraphics[width=0.49\textwidth]{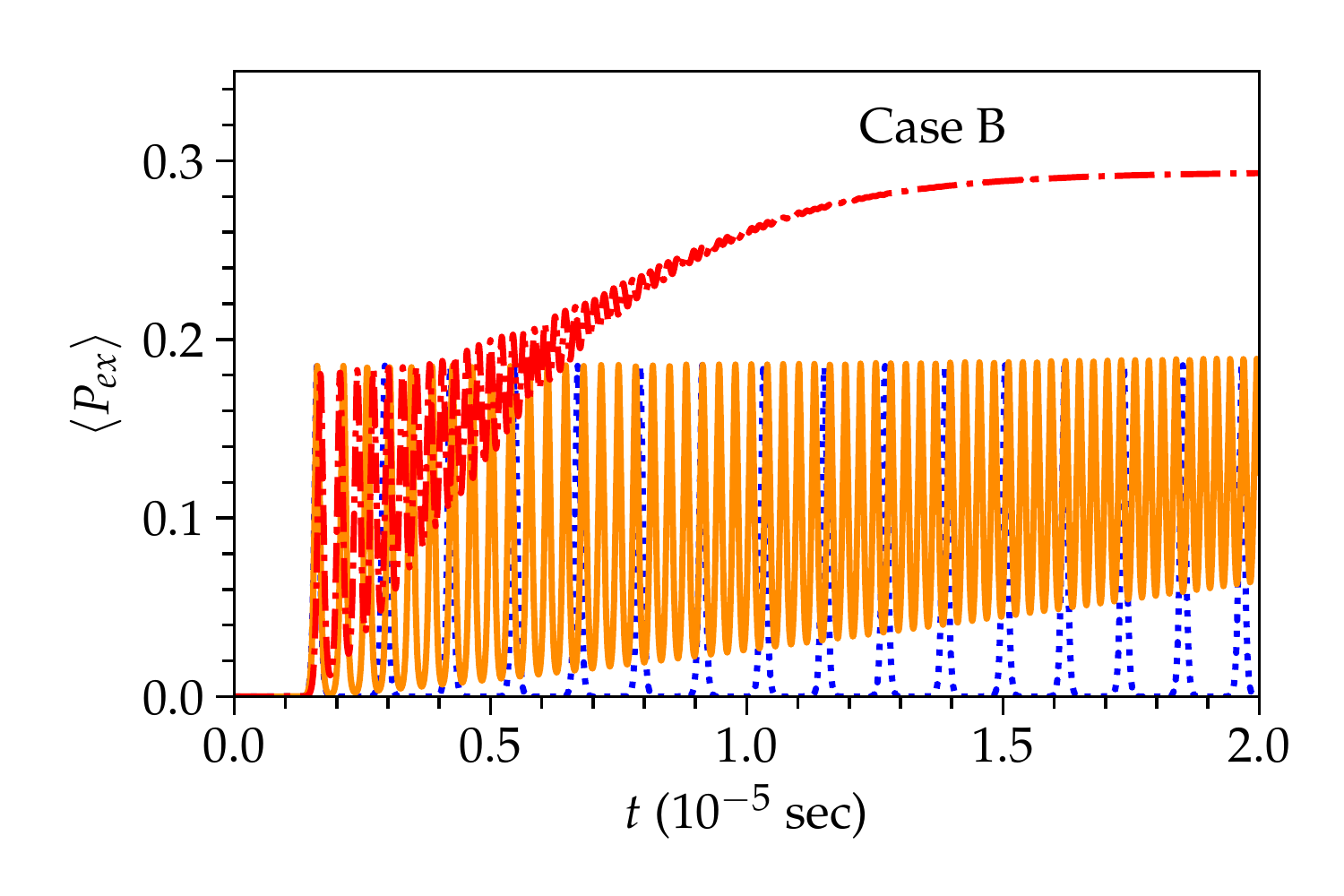}
\includegraphics[width=0.49\textwidth]{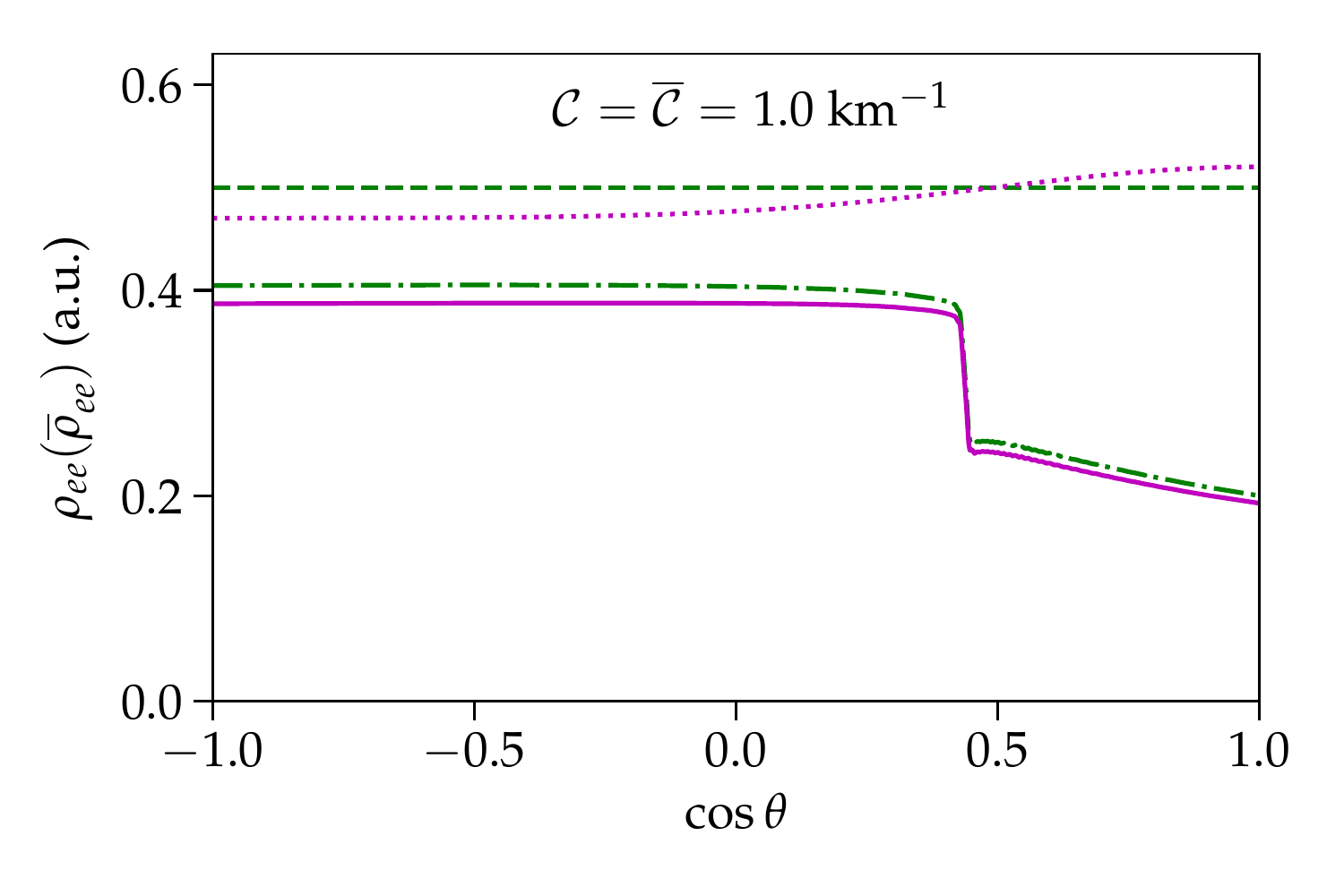}
\includegraphics[width=0.49\textwidth]{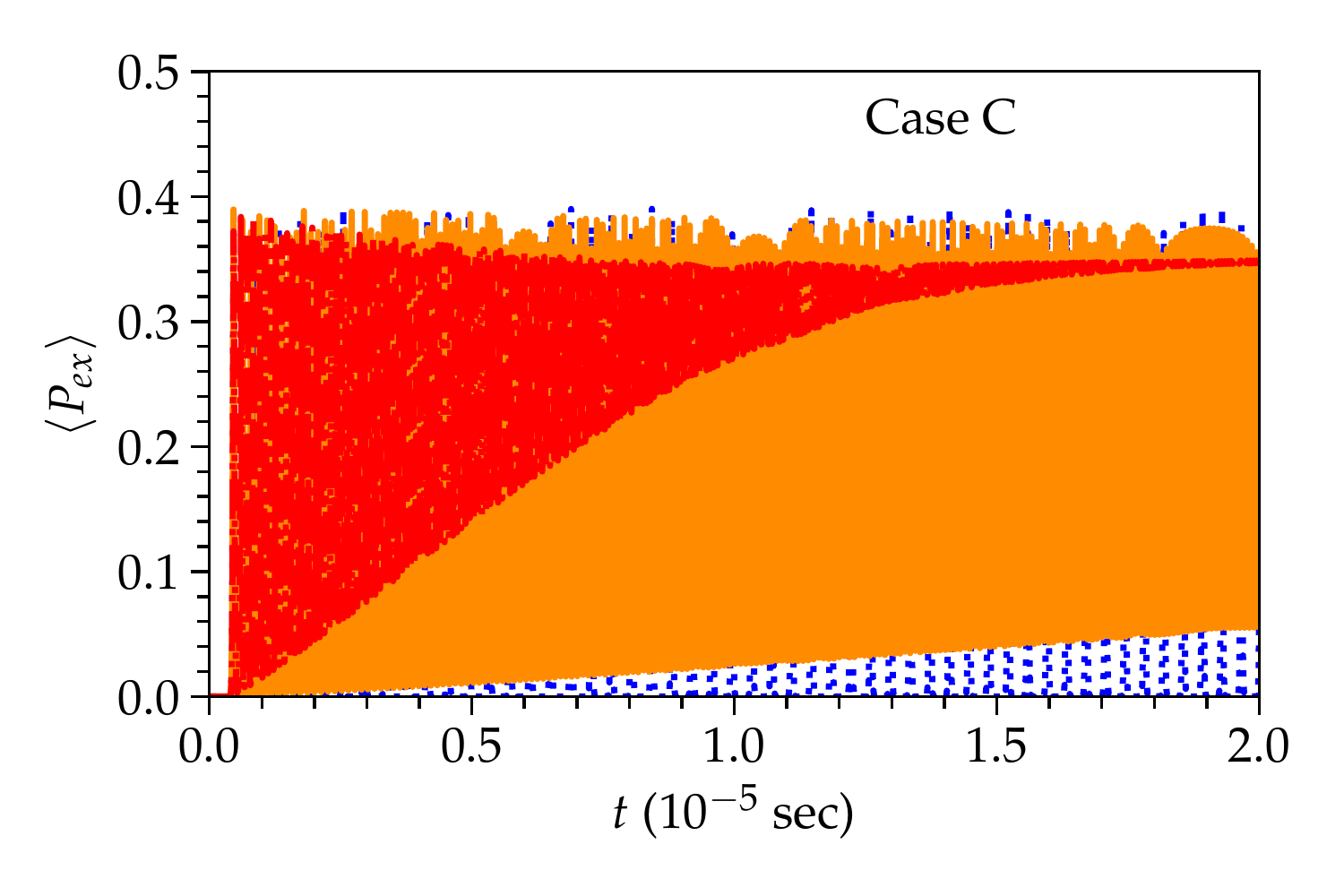}
\includegraphics[width=0.49\textwidth]{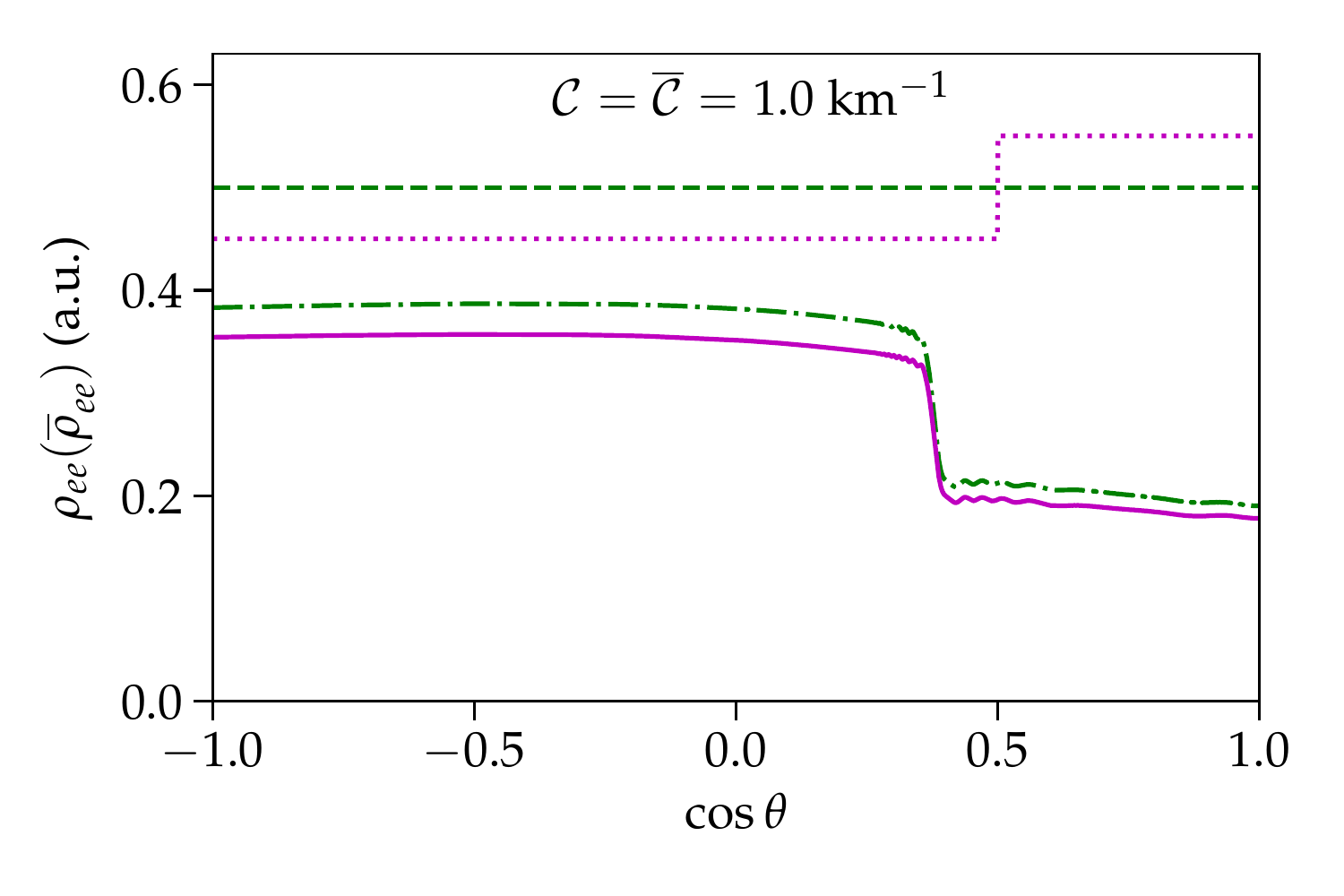}
\caption{{\it Left:} Angle-averaged flavor transition probability for the three Cases described in Eqs.~\ref{cases1}--\ref{cases3} for $E = 50$~MeV and $\Delta m^{2} = 2.5 \times 10^{-6}$ eV$^{2}$, and $\mathcal{C} = \bar{\mathcal{C}}$. {\it Right:} Initial and final ($t=2\times 10^{-5}$~s) angular distributions for $\mathcal{C} = \bar{\mathcal{C}} = 1.0$ km$^{-1}$. As the collision strength increases, flavor conversions are enhanced and a sharp feature appears in the angular distributions of $\nu_e$'s and $\bar\nu_e$'s.}
\label{Fig2}
\end{figure*}

Reference~\cite{Shalgar:2020xns} has shown that flavor conversions can be affected by the vacuum term, especially in the non-linear regime. In this case, finer and less regular structures appear in the angular distributions. Hence, in the numerical results reported below, we fix $\Delta m^{2}=2.5\times 10^{-6}$ eV$^{2}$, which is much smaller than the real value of $\Delta m^{2}$, in order to preserve the periodic structure of flavor mixing~\cite{Shalgar:2020xns}.  We also assume $E=50$~MeV, i.e. $\omega = 1.27 \times 10^{-4}$~km$^{-1}$. 

The simplest case for which we can study the effect of collisions is for $\mathcal{C} = \bar{\mathcal{C}}$. It should be noted that this is not self-consistent with the initial angular distributions of $\nu_e$ and $\bar\nu_e$, i.e.~the fact that the $\bar\nu_e$ angular distribution is forward peaked while the $\nu_e$ one is isotropic implies that the collision term for neutrinos is larger than the one for antineutrinos. However, this benchmark case with $\mathcal{C} = \bar{\mathcal{C}}$ helps to gain intuition on the interplay between flavor conversions and collisions. 

\begin{figure*}
\includegraphics[width=0.49\textwidth]{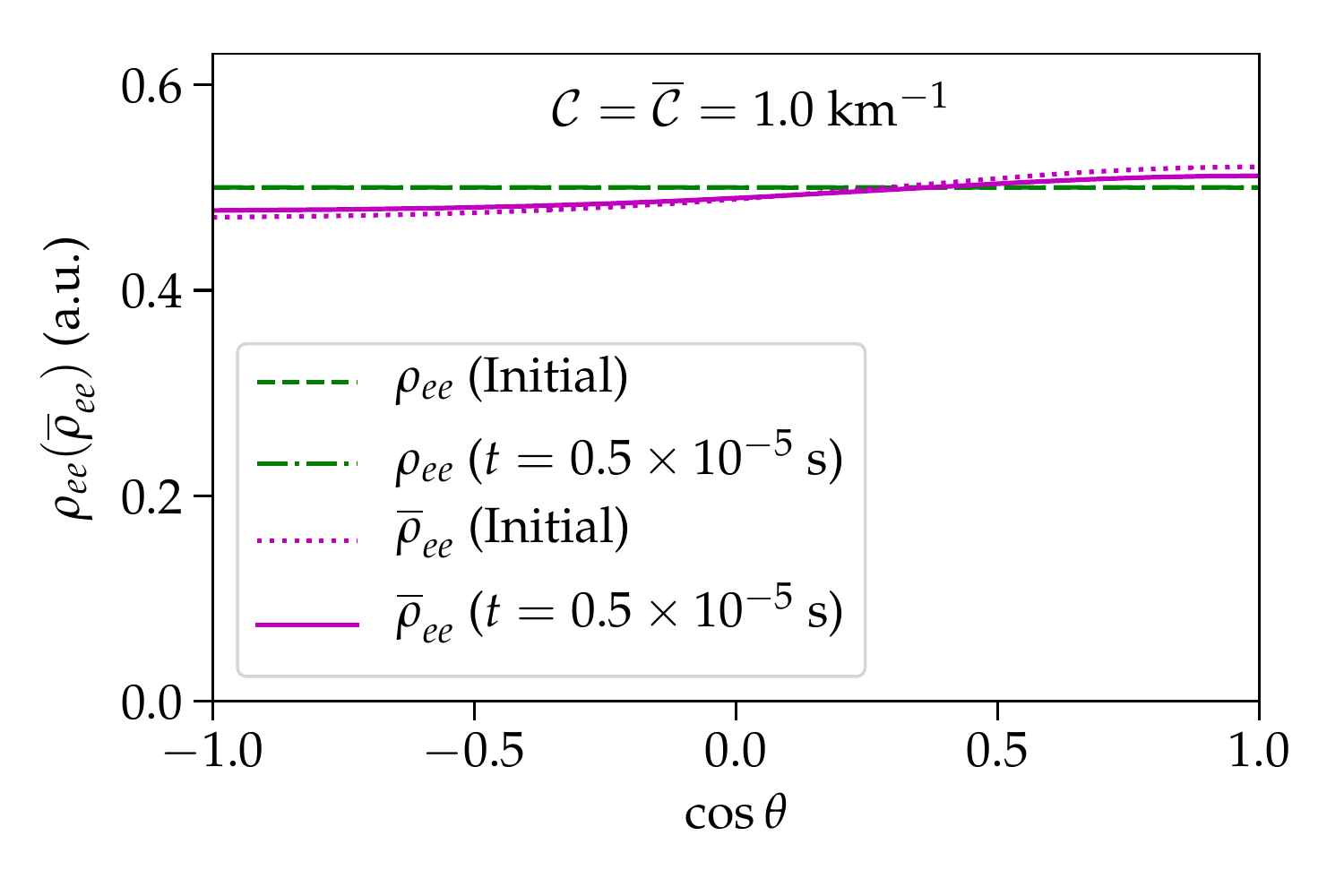}
\includegraphics[width=0.49\textwidth]{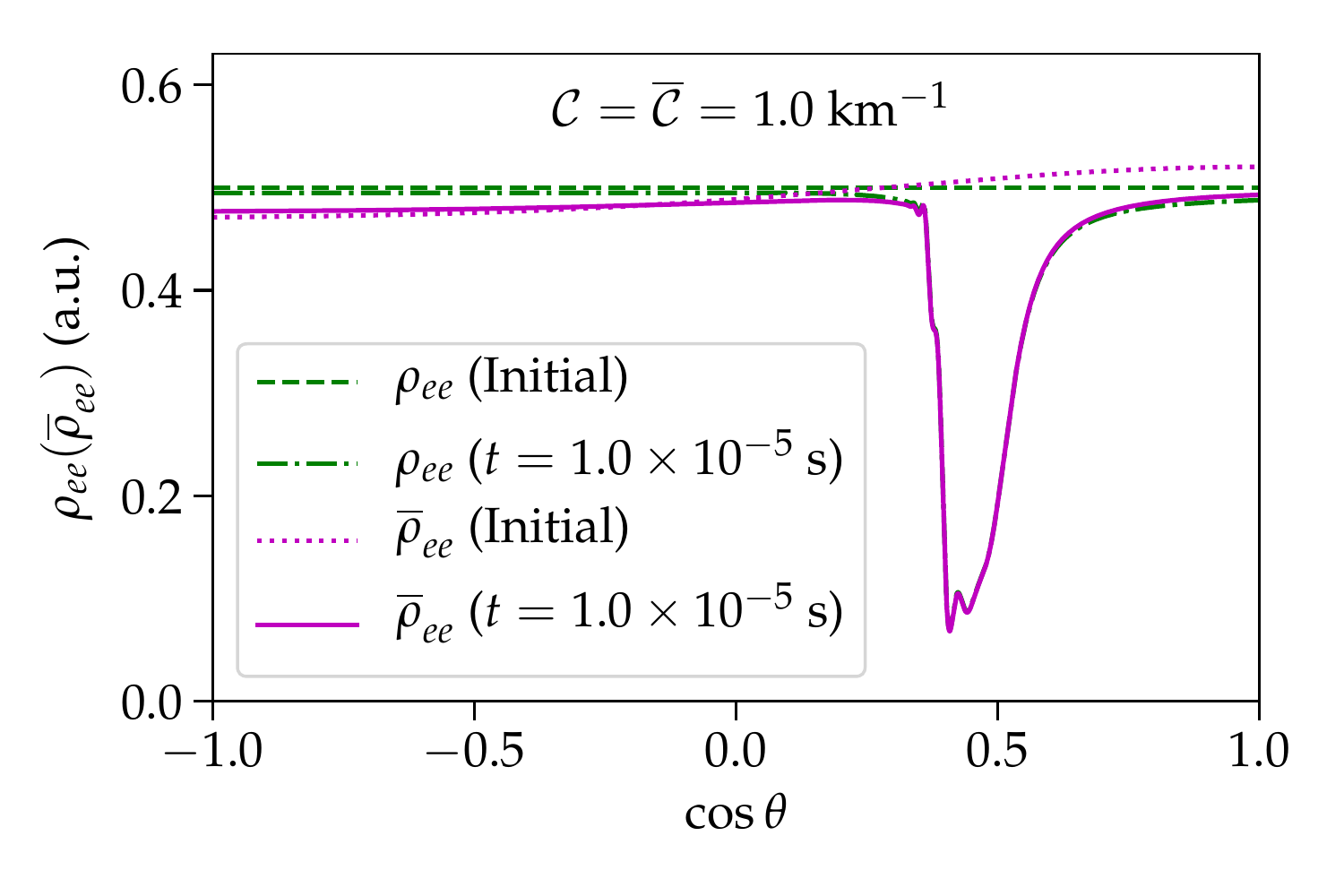}
\includegraphics[width=0.49\textwidth]{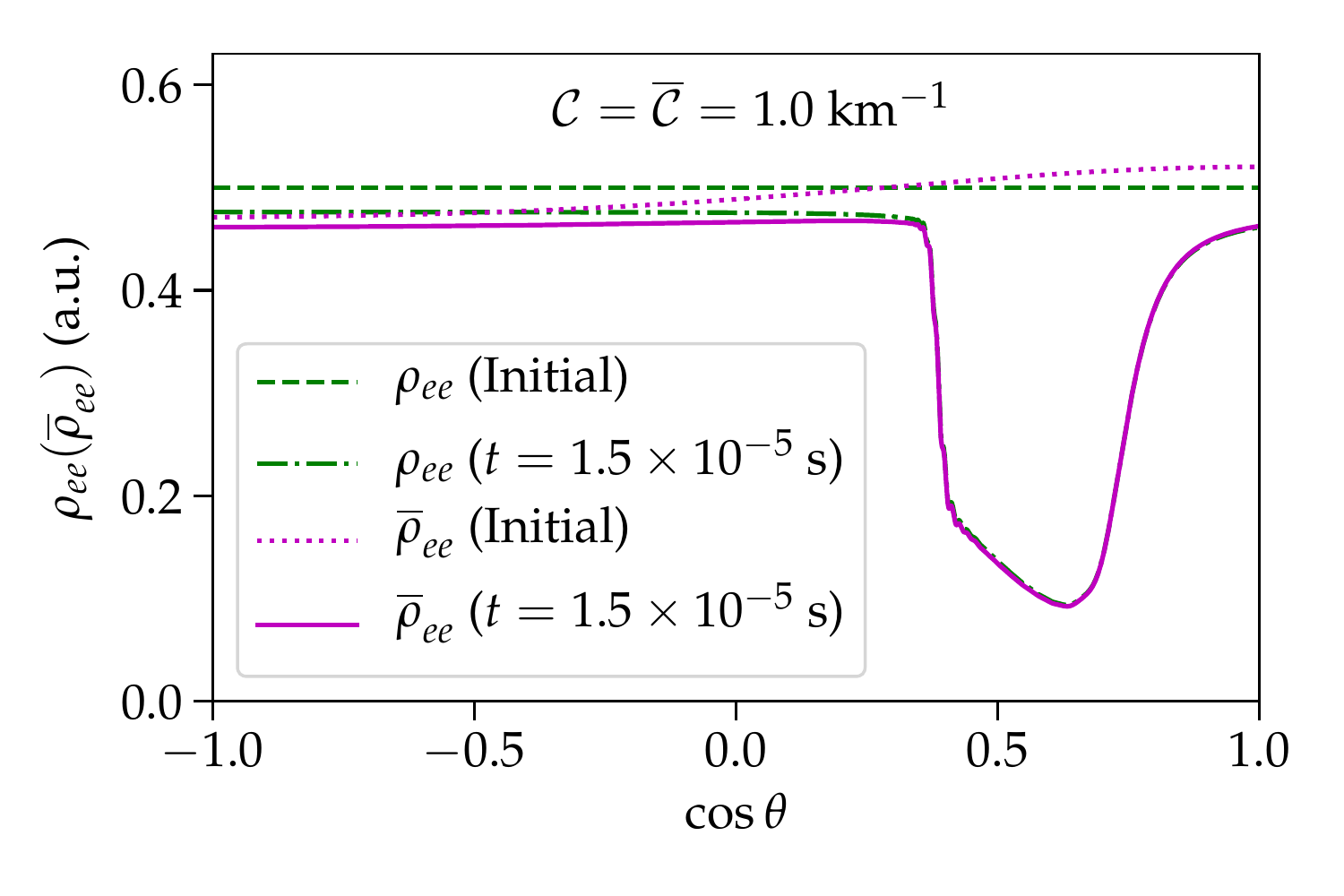}
\includegraphics[width=0.49\textwidth]{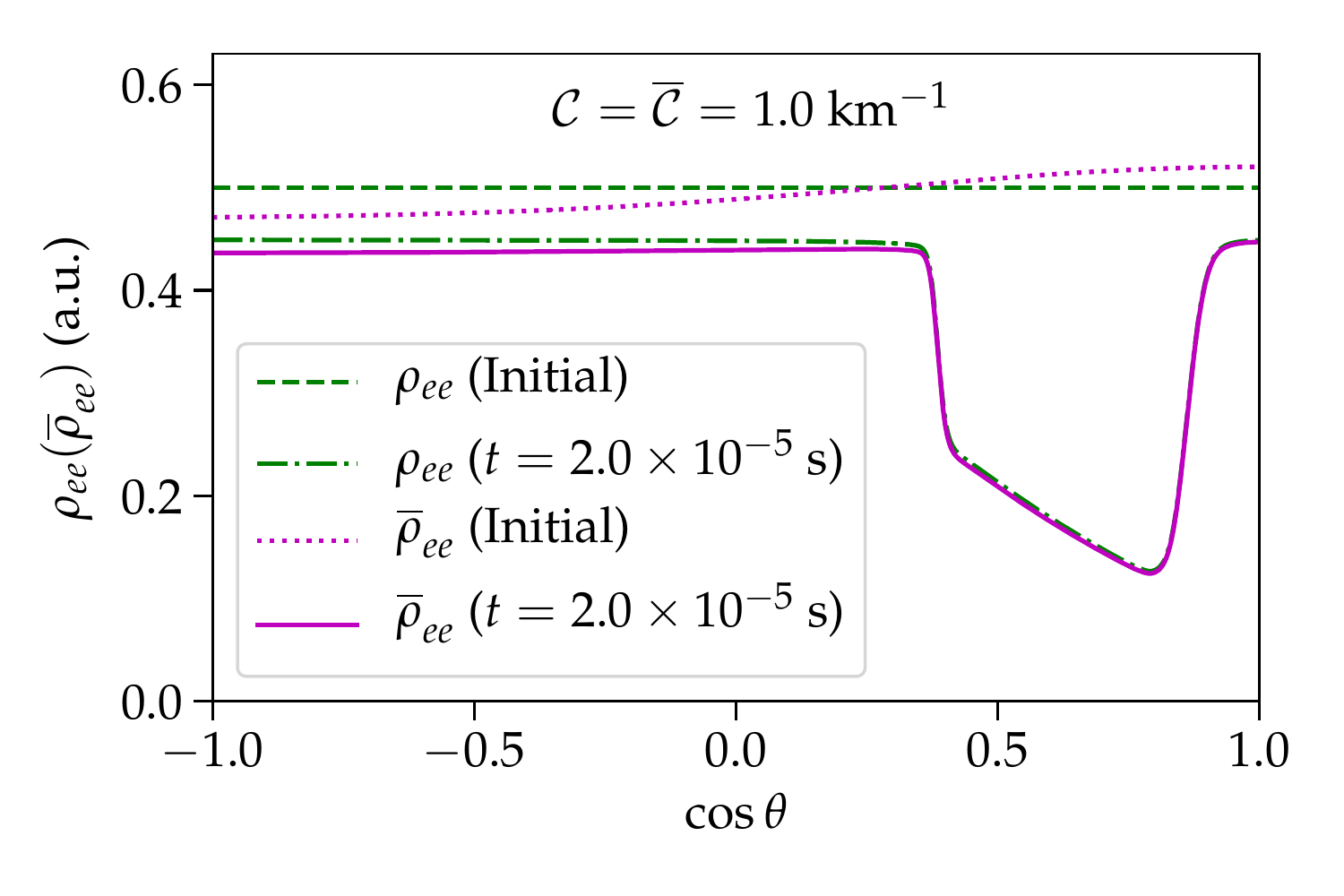}
\caption{Angular distributions of $\nu_{e}$ and $\bar{\nu}_{e}$ at four representative times for Case A and $\mathcal{C} = \bar{\mathcal{C}}=1.0$ km$^{-1}$. All the other parameters are same as the ones used in Fig.~\ref{Fig2}.}
\label{Fig2p}
\end{figure*}

In order to gauge the amount of flavor mixing, we introduce the flavor transition probability averaged over the angles:
\begin{eqnarray}
\label{eq:Pex}
\langle P_{ex} \rangle(t) =1 - \frac{\int \rho_{ee}(\theta,t) d\cos\theta - \int \rho_{xx}(\theta,0) d\cos\theta}
{\int \rho_{ee}(\theta,0) d\cos\theta - \int \rho_{xx}(\theta,0) d\cos\theta}\ .\nonumber\\
\end{eqnarray}

The left panels of Fig.~\ref{Fig2} show the temporal evolution of $\langle P_{ex} \rangle$ for Cases A, B, and C from top to bottom, and for $\mathcal{C} = \bar{\mathcal{C}} = 0, 0.1,$ and $1$~km$^{-1}$. 
One can clearly see that, for Cases A and B, the inclusion of the collision term enhances flavor conversions beyond the maximum value that is otherwise reached in the absence of collisions; however the collision strength is so weak that it does not affect the onset of the non-linear regime. The enhancement of the flavor conversion probability, mostly prominent in Cases A and B, is due to the fact that collisions reshuffle the neutrino densities across angular modes, creating a configuration prone to larger flavor mixng. The right panels of Fig.~\ref{Fig2} show the initial ($t=0$~s) and final ($t=2 \times 10^{-5}$~s) angular distributions, and one can see that the final angular distribution is such that it shows a sharp drop in the same angular  region where the initial flavor distribution has an ELN crossing.
Case C has a slightly different behavior, in the sense that the flavor conversion probability reaches an asymptotic value that is lower than the maximum allowed value that can be reached in the absence of collisions (see bottom left panel), but the time-averaged flavor conversion probability still shows a marginal enhancement. This should be considered as an example of the fact that collisions do not always enhance conversions, but the final outcome depends on the subtle interplay between collisions and neutrino mixing. Moreover, as the strength of collisions increases to very large values (larger than the ones shown in Fig.~\ref{Fig2}), the ELN crossings are erased before flavor conversions set in. 

The trend displayed in Fig.~\ref{Fig2} can be better grasped by looking at the snapshots of the temporal evolution for  Case A in Fig.~\ref{Fig2p} and the animations provided as \href{https://sid.erda.dk/share_redirect/Cb7Xyrai4n/index.html}{Supplemental Material}; the latter show the temporal evolution of the $\nu_e$ and $\bar\nu_e$ angular distributions is shown together with the temporal evolution of the polarization vectors, $\vec{P}$ and $\vec{\bar{P}}$ defined as
\begin{eqnarray}
P_{a} = \textrm{Tr}(\rho \sigma_{a}) \quad \mathrm{and} \quad \bar{P}_{a} = \textrm{Tr}(\bar{\rho} \sigma_{a})\ ,
\end{eqnarray}
where $a=x,y,z$ and $\sigma_{a}$ are the Pauli  matrices. From the movies, we can see that we start with the polarization vectors of neutrinos and antineutrinos that are aligned to the $z$ direction; however, while in the absence of collisions they would keep precessing around the $z$ axis~\cite{Shalgar:2020xns}, collisions are responsible for pushing the polarization vectors away from their original direction. The temporal evolution is such that the polarization vectors (and the angular distributions) do not go back to their original configurations, as it would happen in the absence of collisions. 

At first, because of collisions, the difference between the $\nu_e$ and $\bar\nu_e$ distributions tends to decrease and the two distributions tend to become similar to each other. This, in turn, is responsible for creating conditions that trigger fast pairwise conversions and the system tends to evolve towards a configuration where the two angular distributions are similar to each other, while developing a sharp feature in the angular distributions that is similar to the spectral split found in the energy distributions for slow $\nu$--$\nu$ conversions~\cite{Duan:2007bt,Fogli:2007bk,Fogli:2008pt}. 

In other words, in the absence of collisions, the total neutrino number density at each angle is conserved, but the inclusion of collisions lifts this restriction. This allows for a relaxation of the system to a natural steady-state, if it exists.

\begin{figure*}
\includegraphics[width=0.49\textwidth]{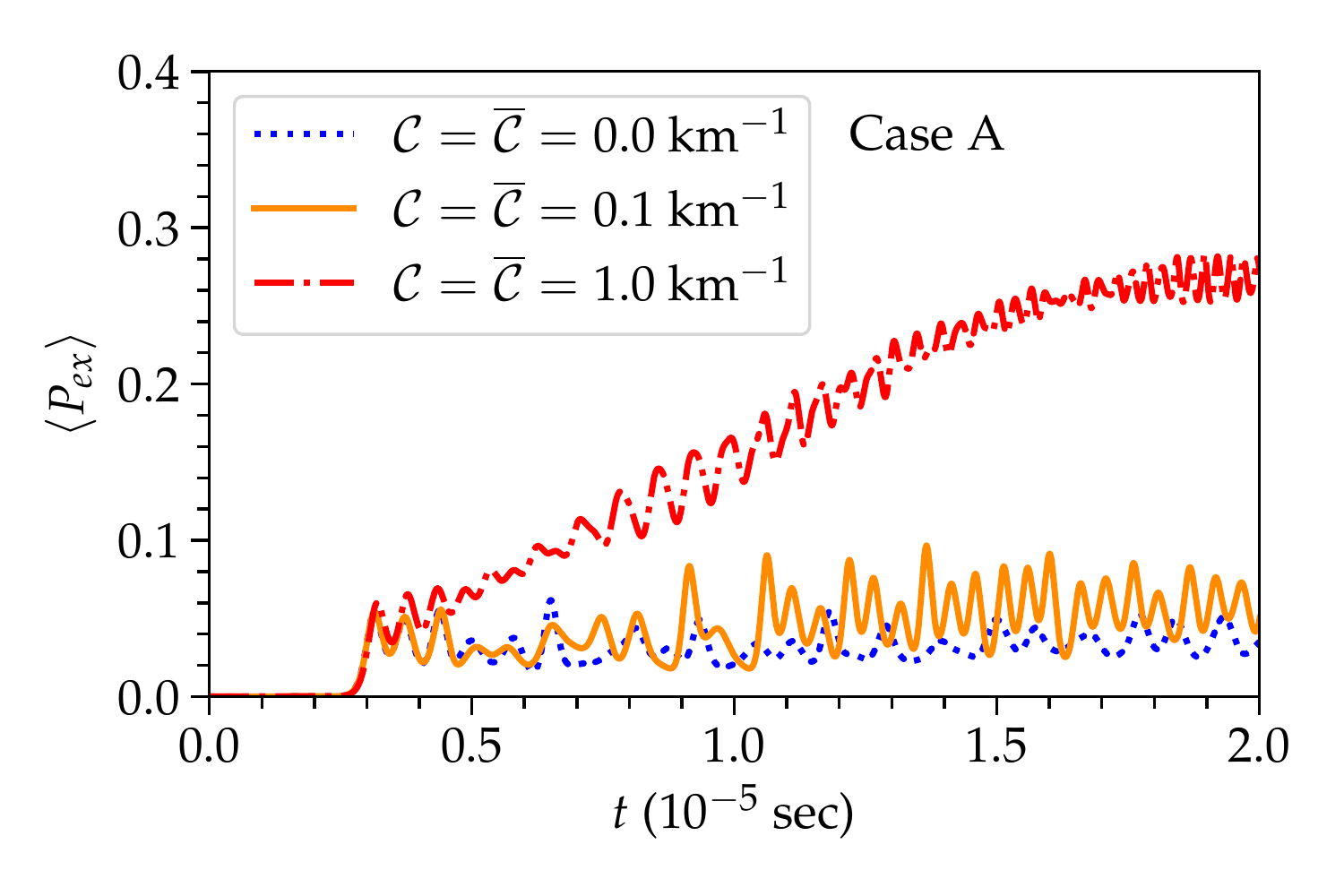}
\includegraphics[width=0.49\textwidth]{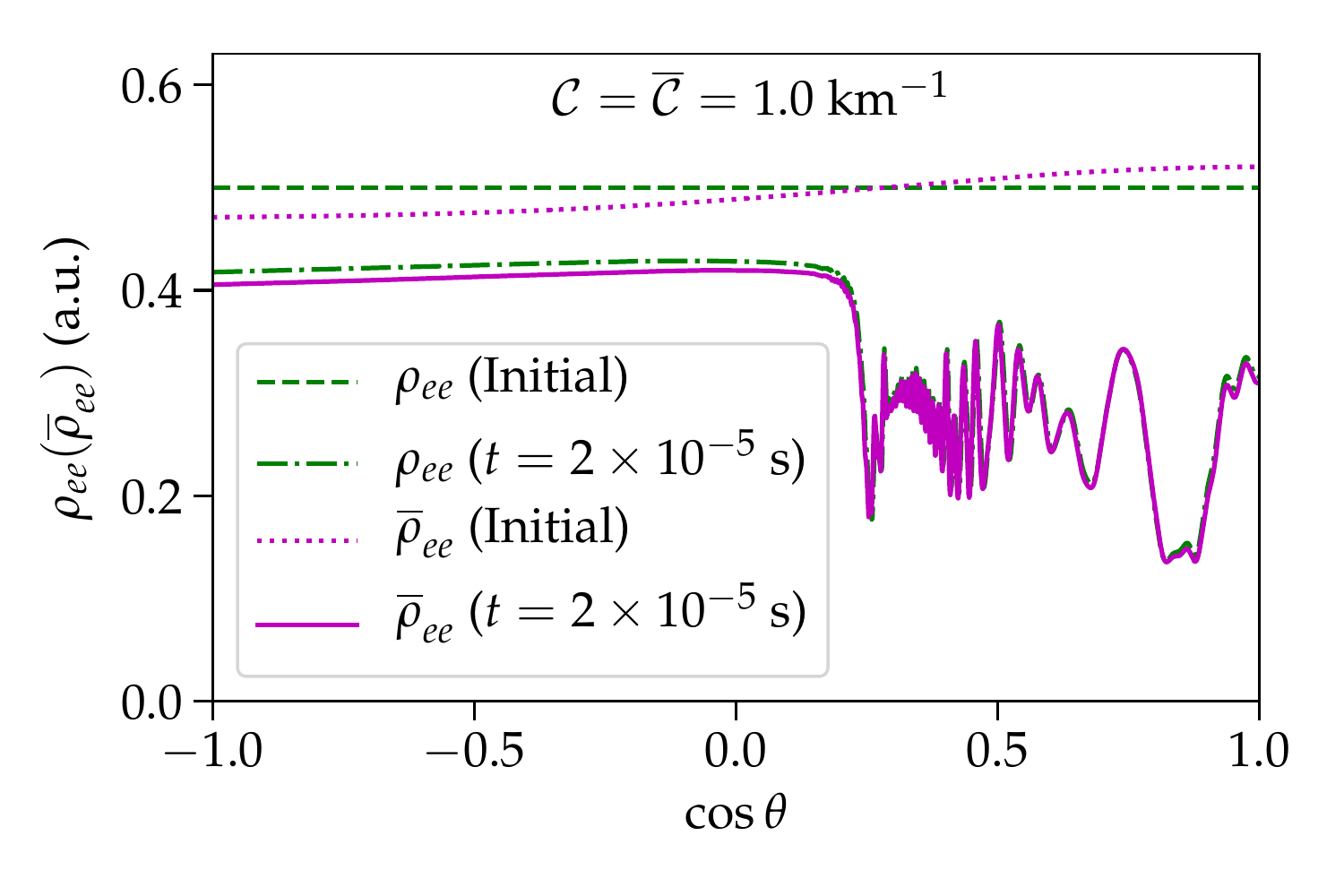}
\includegraphics[width=0.49\textwidth]{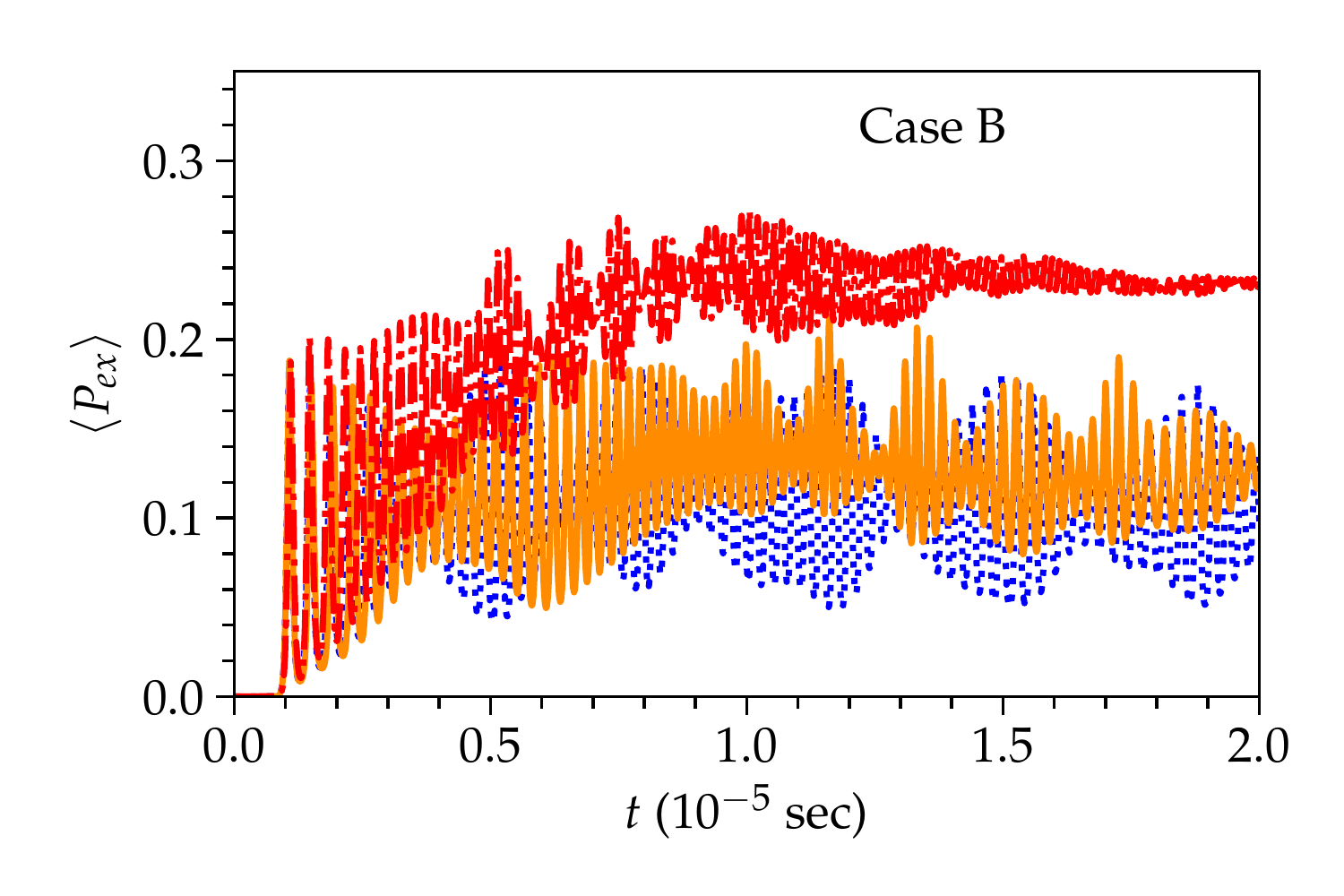}
\includegraphics[width=0.49\textwidth]{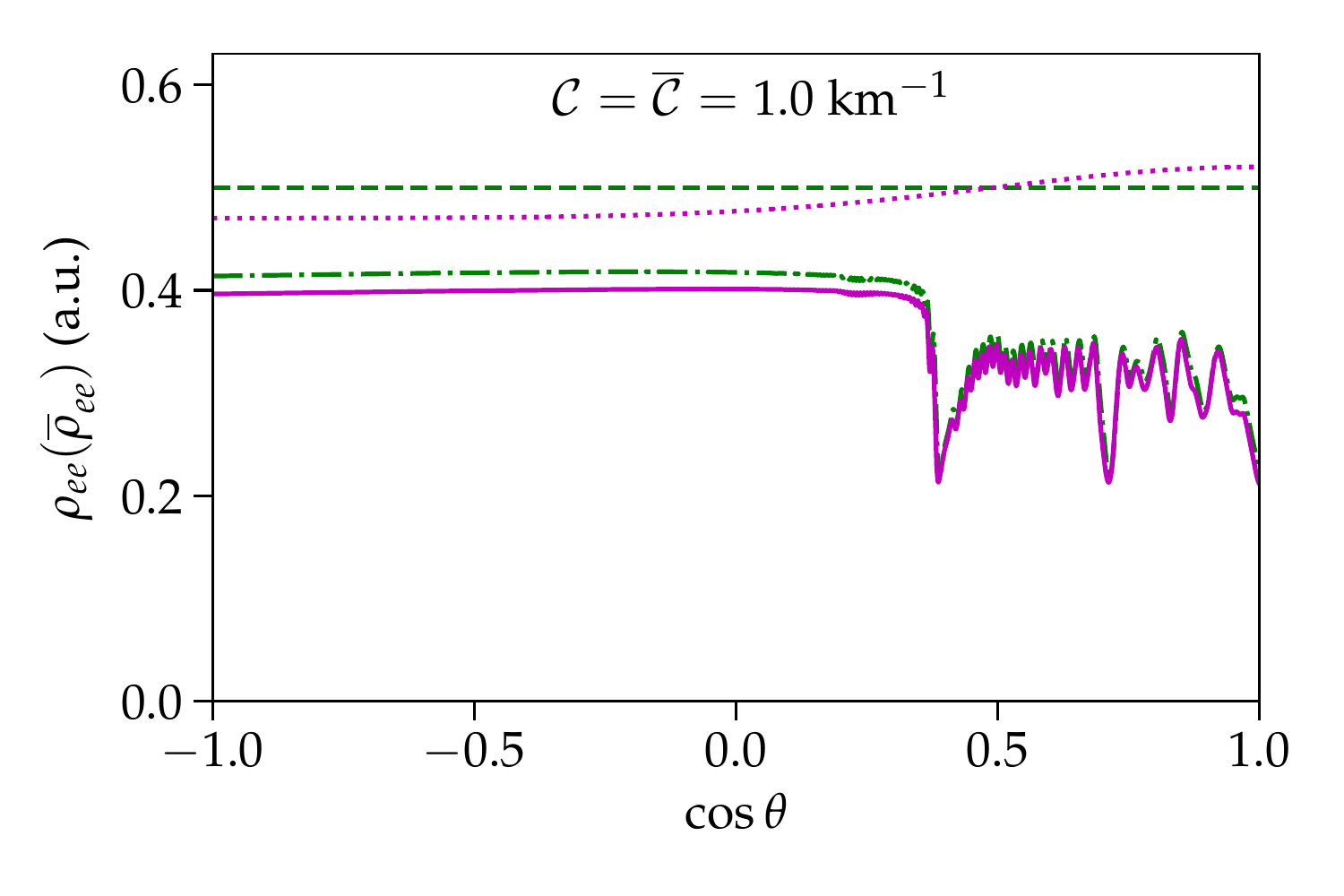}
\includegraphics[width=0.49\textwidth]{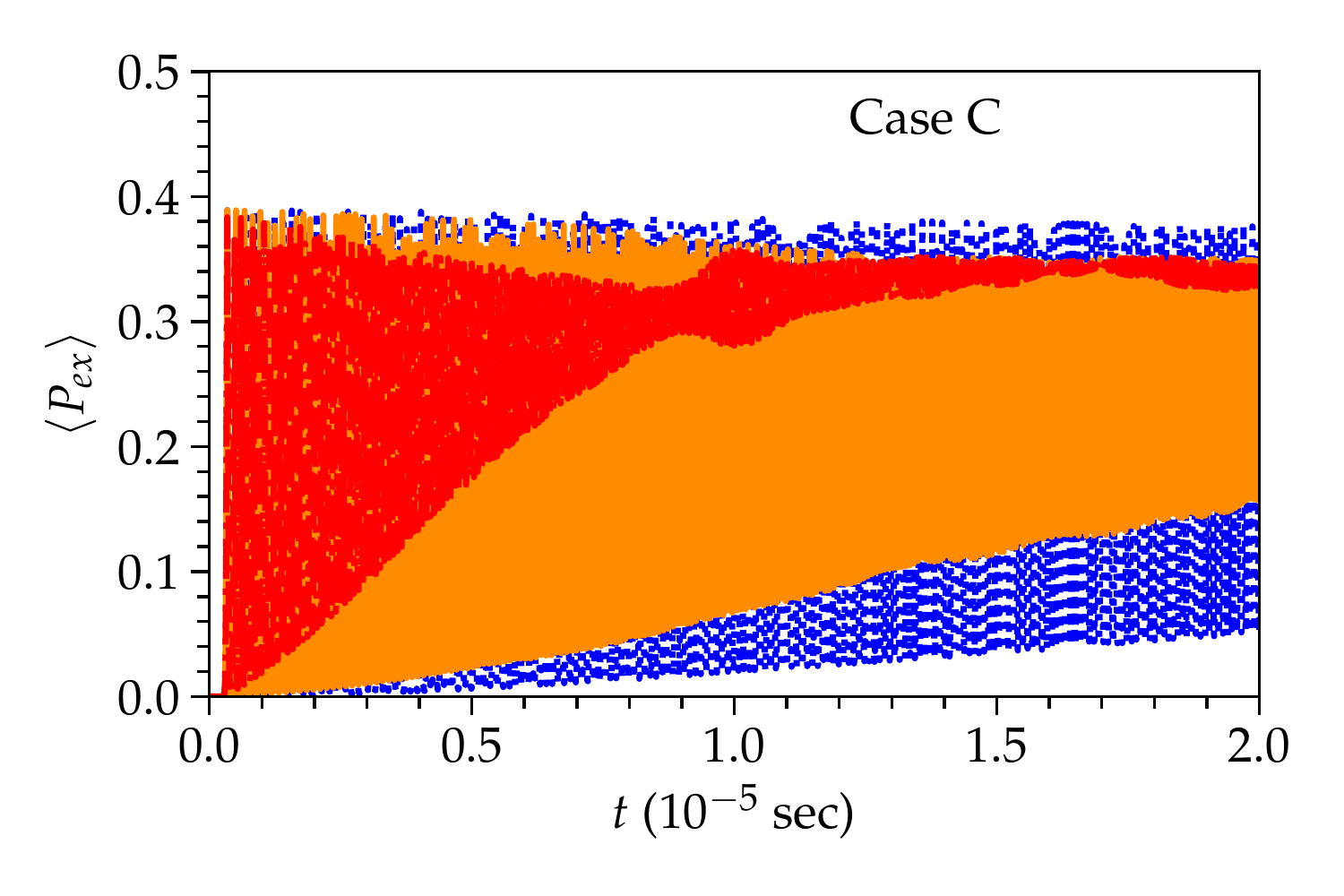}
\includegraphics[width=0.49\textwidth]{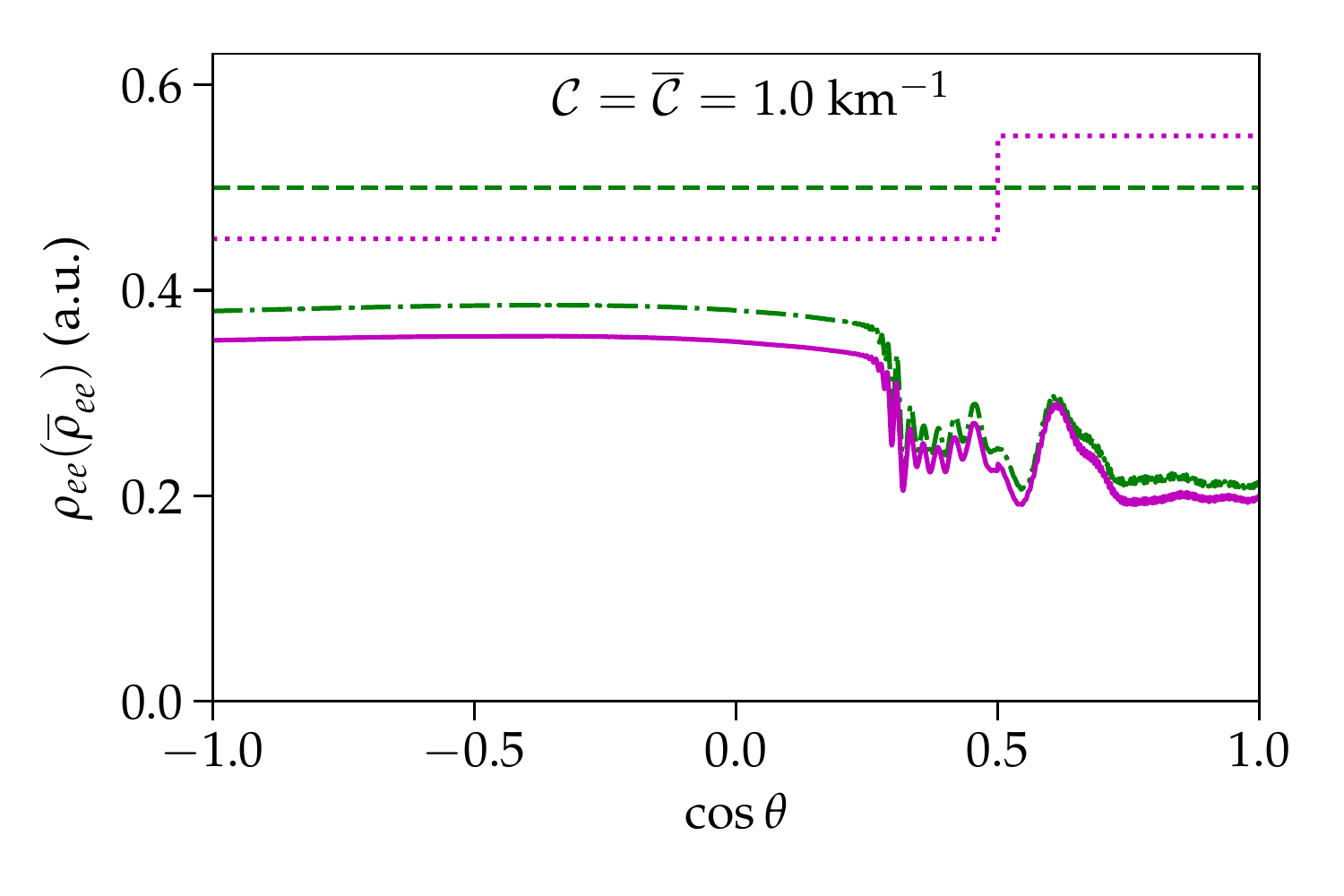}
\caption{Same as Fig.~\ref{Fig2} but for $\Delta m^{2} = 2.5 \times 10^{-3}$ eV$^{2}$. The increase in $\Delta m^2$ is responsible for the appearance of small scale structures in the transition probability as well as in the angular distributions.}
\label{Fig3}
\end{figure*}

The limit of small vacuum frequency is useful to study due to the lack of small scale angular or temporal structures. However, it is far from what one would expect in a realistic scenario. 
One of the key effects of a realistic value of the vacuum frequency is the emergence of small scale angular structures~\cite{Shalgar:2020xns}. 
This is clearly evident in  Fig.~\ref{Fig3}, where  $\Delta m^{2} = 2.5 \times 10^{-3}$~eV$^2$ ($\omega = 0.127$ km$^{-1}$) makes the picture more complex. However,  the overall tendency of the system to achieve more flavor conversions in the presence of collisions remains. 
Similar to Fig.~\ref{Fig2}, the enhancement of flavor conversions in Fig.~\ref{Fig3} is much more noticeable in Cases A and B, as opposed Case C. In fact, for Case C, significant flavor conversion is possible even in the absence of collisions. This leaves less scope for collisions to enhance flavor conversions. 

\subsection{Different collision strength for neutrinos and antineutrinos}
\begin{figure*}
\includegraphics[width=0.49\textwidth]{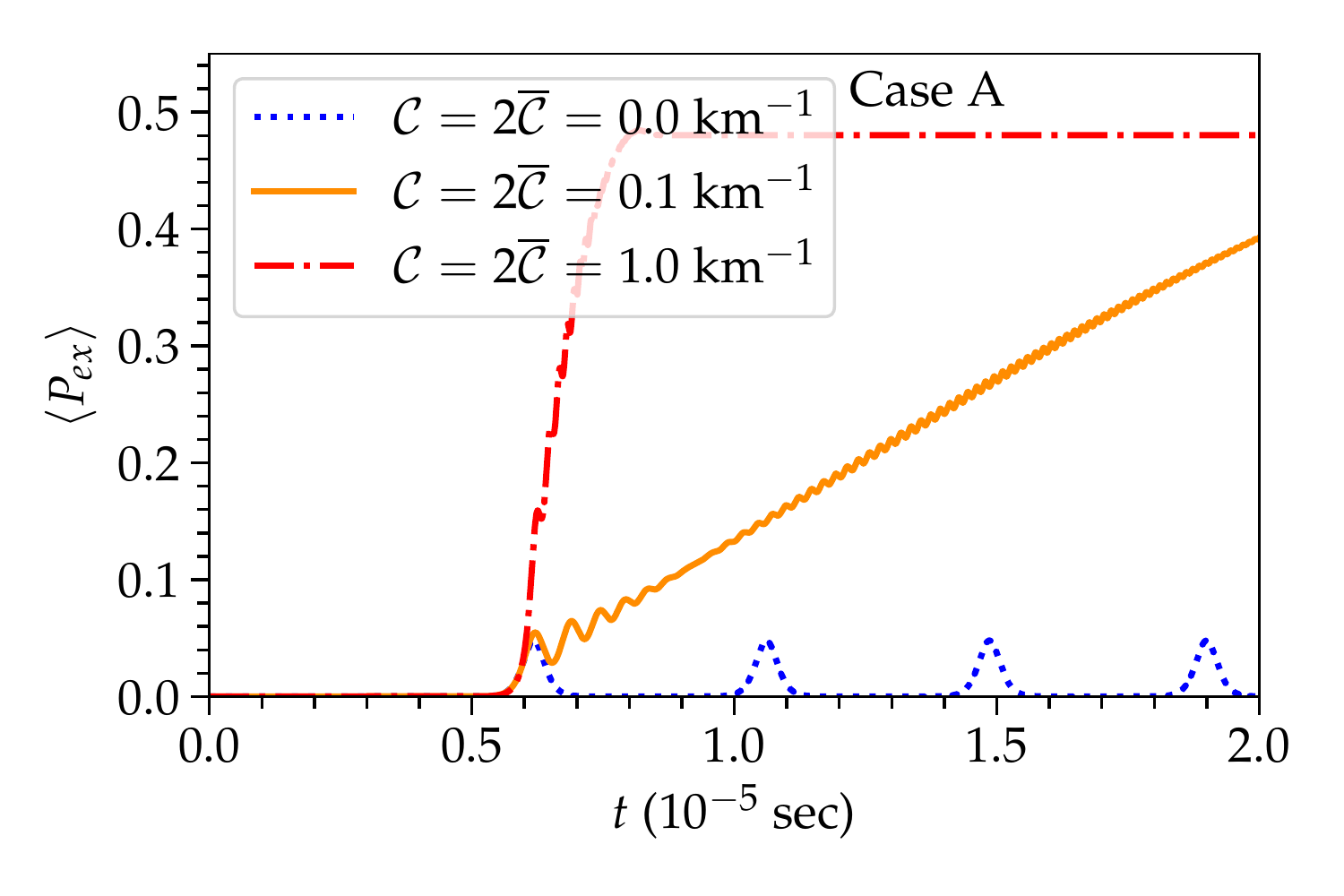}
\includegraphics[width=0.49\textwidth]{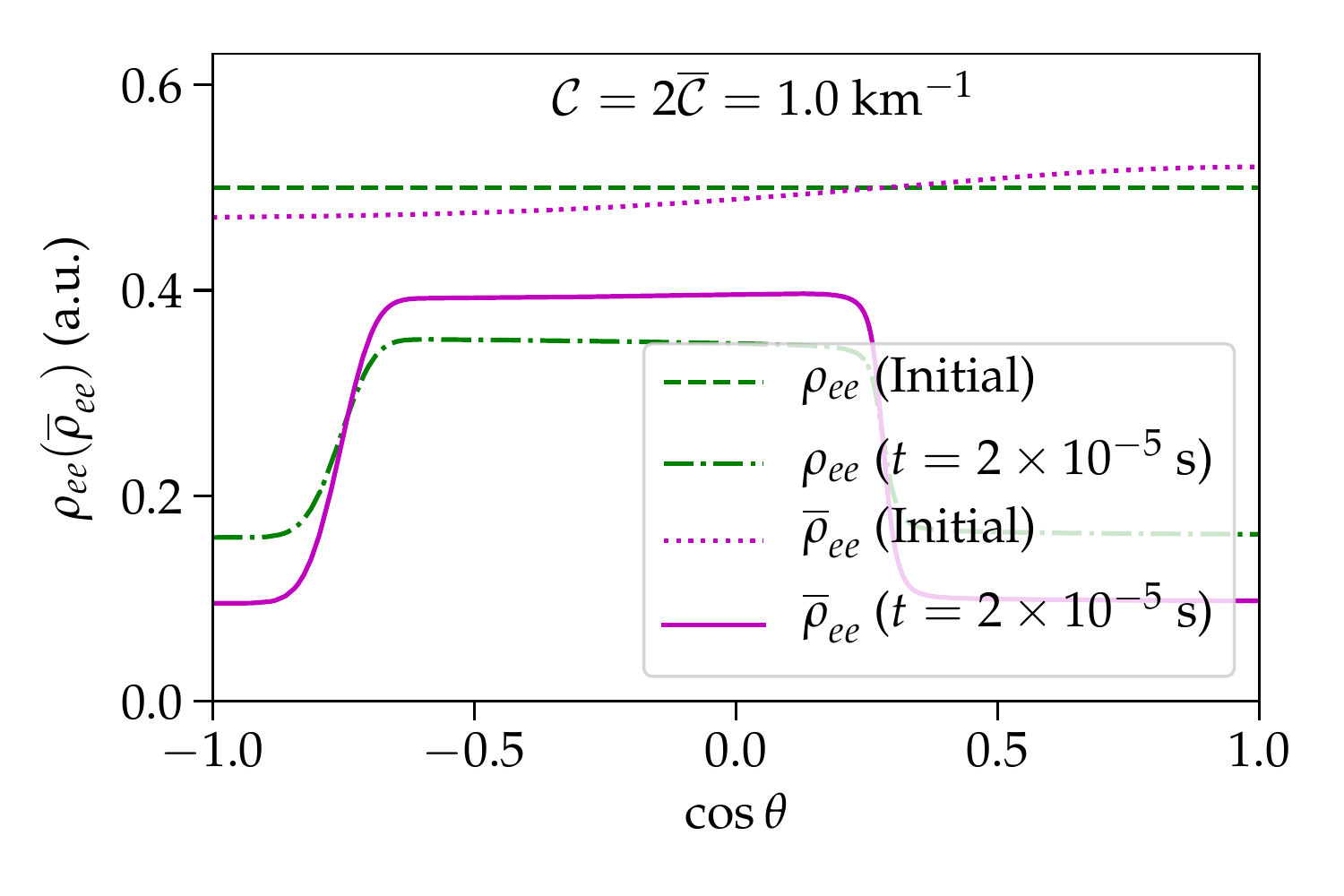}
\includegraphics[width=0.49\textwidth]{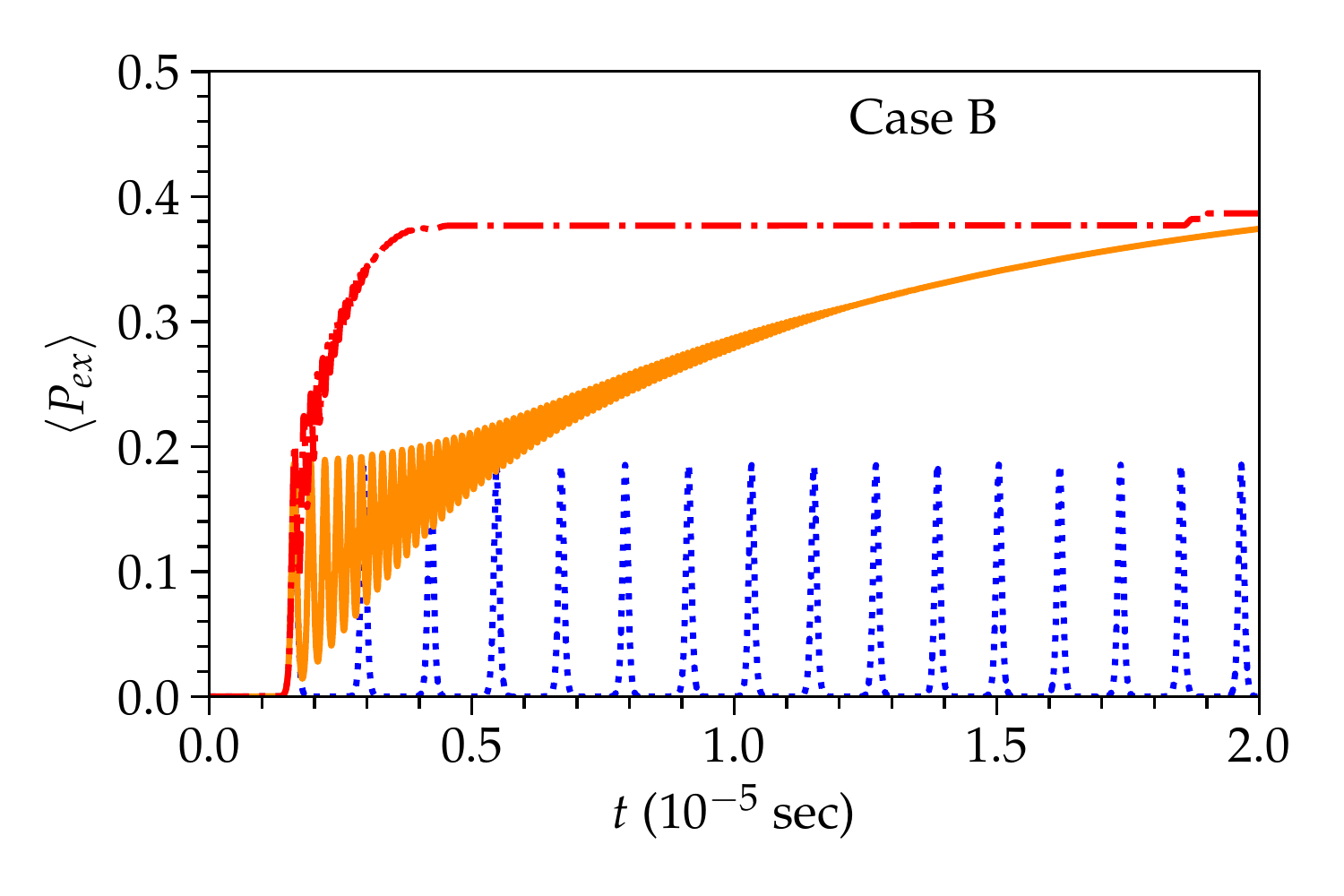}
\includegraphics[width=0.49\textwidth]{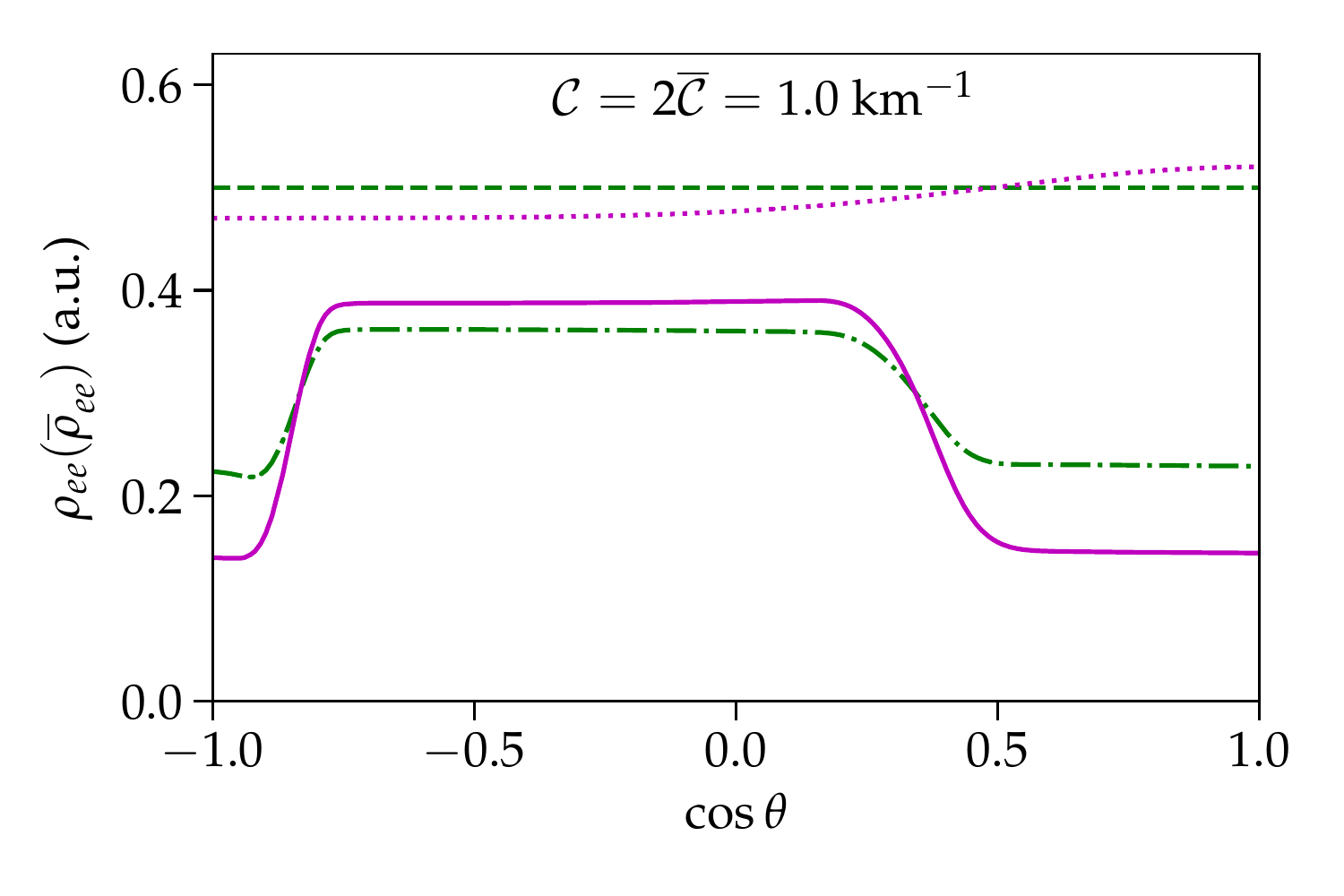}
\includegraphics[width=0.49\textwidth]{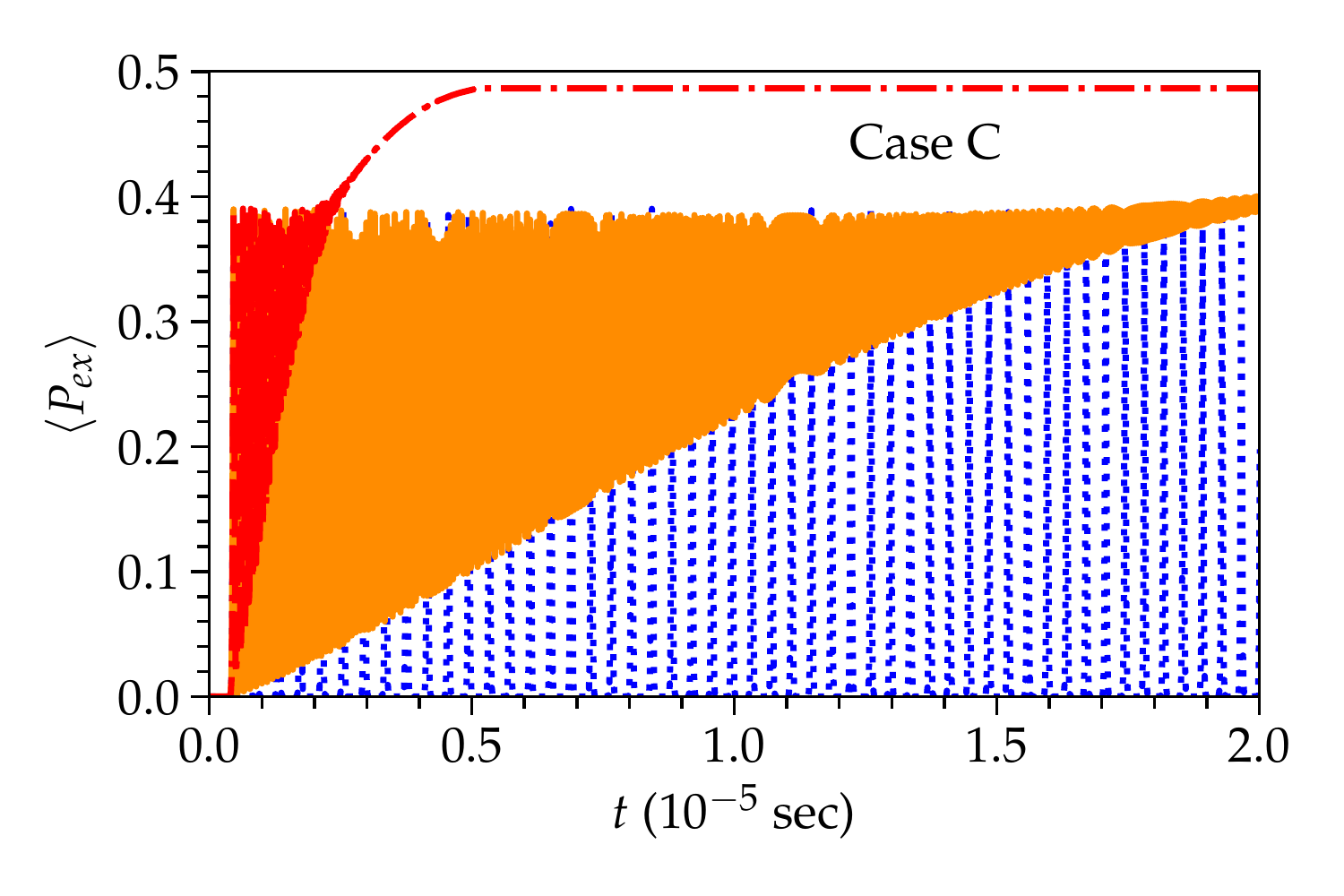}
\includegraphics[width=0.49\textwidth]{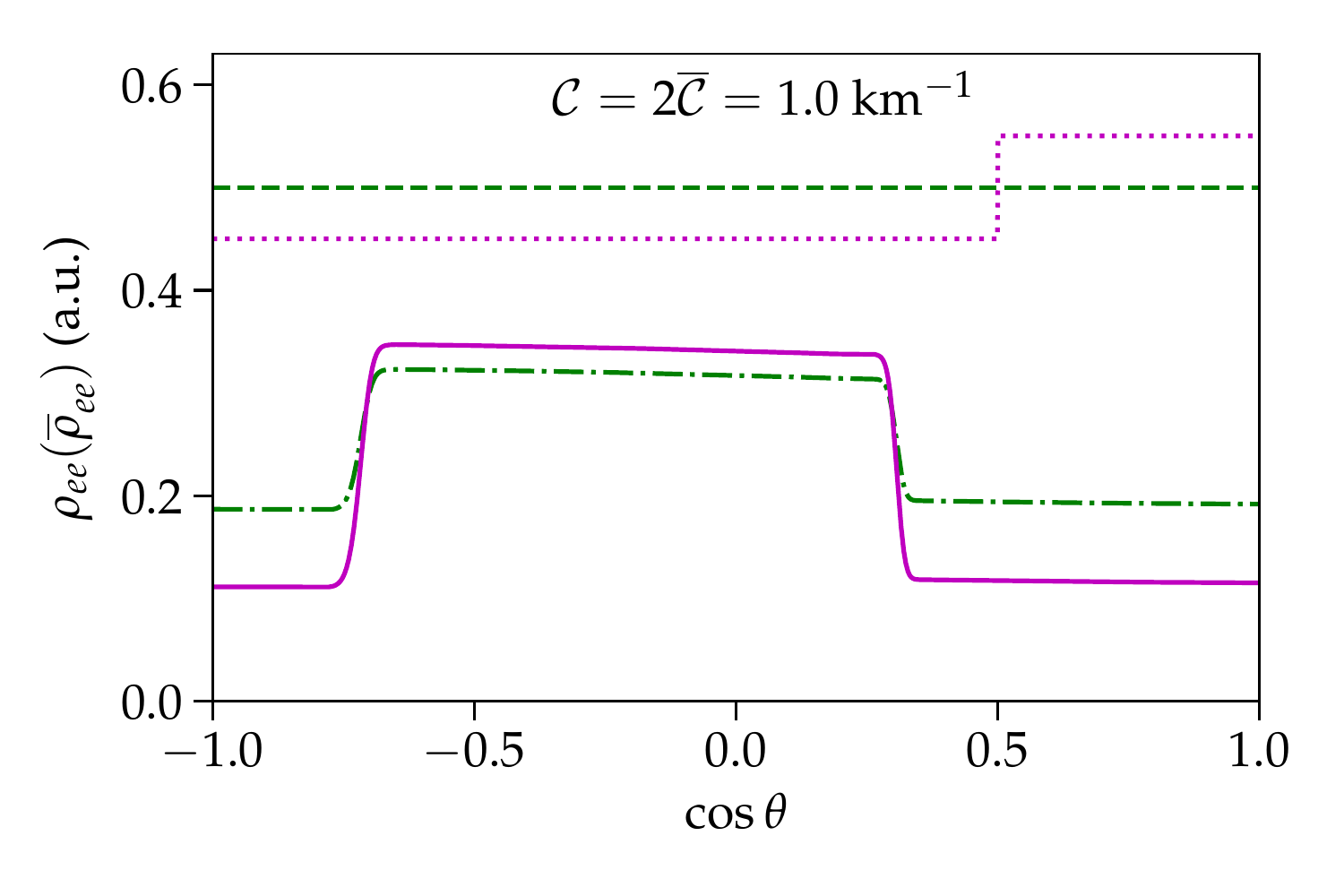}
\caption{Same as Fig.~\ref{Fig2} but for $\mathcal{C} = 2\bar{\mathcal{C}}$. The different collision term for neutrinos and antineutrinos fosters the development of a second sharp feature in the angular distributions (see Fig.~\ref{Fig2} for comparison) as well as an enhancement of flavor mixing.}
\label{Fig4}
\end{figure*}

The study of flavor evolution for $\mathcal{C}=\bar{\mathcal{C}}$ is illuminating, but it is not expected to be present in a real astrophysical scenario, where $\mathcal{C} > \bar{\mathcal{C}}$.
Figure~\ref{Fig4} is the analogous of Fig.~\ref{Fig2} but for $\mathcal{C} = 2 \bar{\mathcal{C}}$. 
The different collision strength for neutrinos and antineutrinos further enhances the transition probability, but the overall physics picture is unchanged. Intriguingly, the different collisional strength between neutrinos and antineutrinos is responsible for the development of a second sharp feature in the angular distributions at $t = 2 \times 10^{-5}$~s (see right panels of Fig.~\ref{Fig4}), in addition to the one developing in correspondence of the initial ELN crossing as also discussed in Fig.~\ref{Fig2}.

A trend is seen in all the results presented in this paper. In the collective regime, all (anti)neutrino momentum modes evolve in sync in the absence of collisions. Collisions break the collectiveness of the evolution and, as more avenues for breaking the collective behaviour are included, there is an enhancement in the conversion probability of neutrinos. This is also evident in the animations of the evolution of the polarization vectors provided as \href{https://sid.erda.dk/share_redirect/Cb7Xyrai4n/index.html}{Supplemental Material}. 

\section{Discussion and outlook}
\label{sec:conc}
In this work, we explore the interplay between fast pairwise conversions and collisions. Ours is among the first attempts to explore the intricate feedback of collisions on the fast flavor conversion physics and the related dynamical change of the (anti)neutrino angular distributions due to collisions. As such, there are several limitations in our approach. Some are inherent to the formalism used in this paper, while others are deliberate simplifications which are instrumental to ensure that our results are not cluttered by too many variables.

We focus on the simplest possible scenario which consists of an angle-independent collision term, whose main function is to change the (anti)neutrino direction, and treat the strength of the collision term and the shape of the angular distribution as independent of each other. 
This assumption is made for purely pedagogical purposes and can be easily relaxed by promoting the collision term anti-commutator as shown in~\cite{Sigl:1992fn,Vlasenko:2013fja,Vlasenko:2014bva,Blaschke:2016xxt}.

We assume that the neutrino number density is conserved by the collision term. This is true only in the case of elastic scattering of neutrinos, but collisions may also absorb or emit neutrinos, most notably though beta and inverse beta reactions~\cite{Bruenn:1985en}. 
 Moreover, we do not let collisions set the initial angular distributions, which is an approximation since the angular distributions of neutrinos are shaped by the collision term, probably simultaneously to the development of flavor conversions. However, including this feedback self-consistently in the calculations increases the computational complexity of the problem dramatically, while our focus is to consider the simplest possible setup to explore the phenomenology of this interplay. 
 
 The feedback effect of collisions on the angular distributions is not as simple as one might imagine at first. Within our simplified setup, we show that the neutrino flavor evolution is significantly affected by the collision term, and fast conversions can be enhanced in the presence of collisions, contrary to naive expectations.

\begin{figure}
\includegraphics[width=0.49\textwidth]{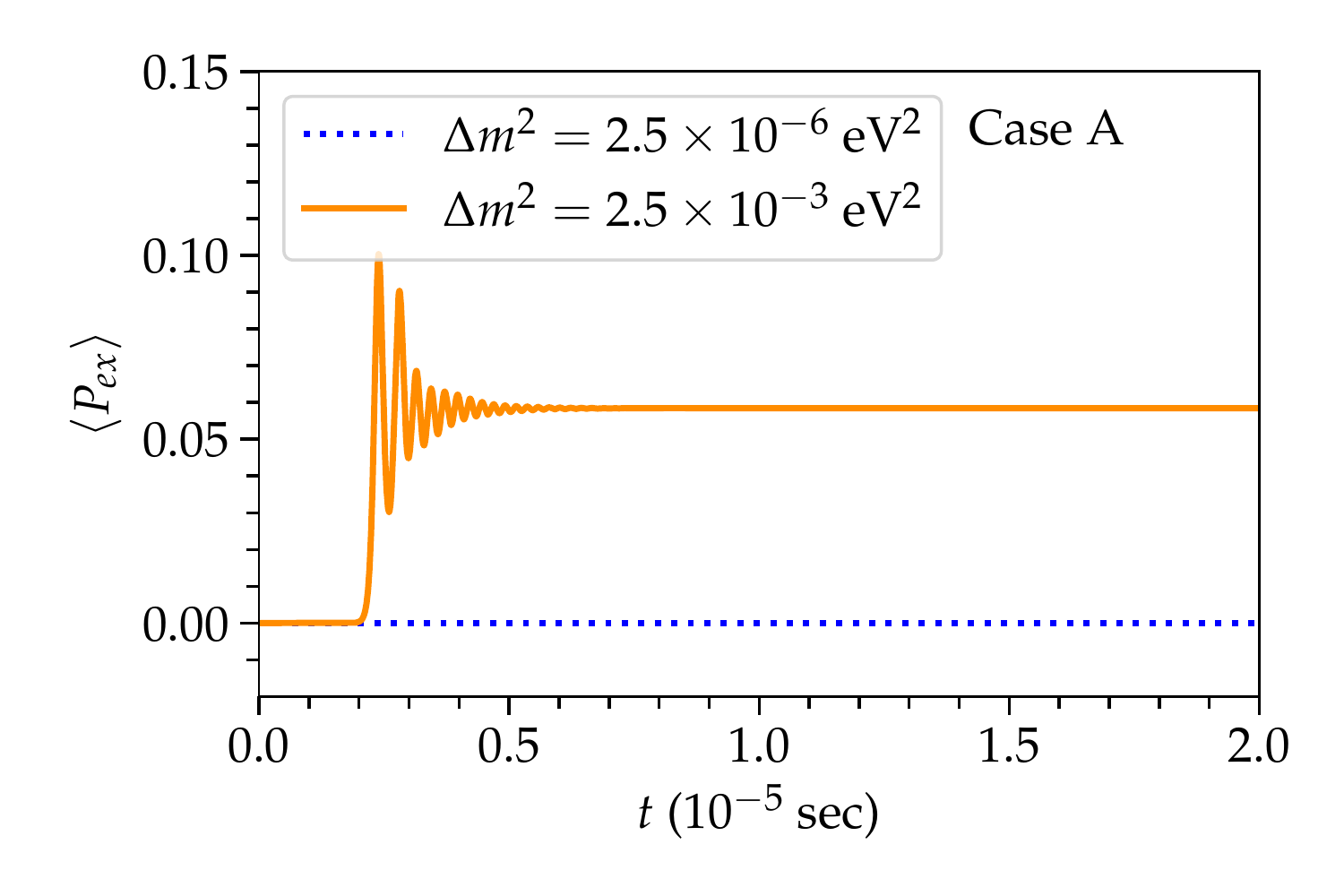}
\caption{Evolution of the angle-averaged flavor transition probability as a function of time for $\mathcal{C} = \bar{\mathcal{C}}$ = 10 km$^{-1}$. A spurious dependence on the initial conditions occurs in the limit of large collision strength. 
}
\label{Fig5}
\end{figure}

When the collision strength is much larger than what we have considered in Sec.~\ref{sec:results}, the ELN crossings can be erased before the system reaches the non-linear regime. Whether this happens or not depends on the value of the initial perturbation that is provided by the vacuum frequency. As shown in Fig.~\ref{Fig5}, collisions damp flavor conversions for $\Delta m^{2} = 2.5 \times 10^{-6}$ eV$^{2}$, but the suppression is less effective for $\Delta m^{2} = 2.5 \times 10^{-3}$ eV$^{2}$. The different trend is due to the difference in the time required for the system to approach the non-linear regime.  In this case, the collision strength is such that the onset time of the non-linear regime depends on $\mathcal{C} (\bar{\mathcal{C}})$.

We  restrict our analysis to the single energy case. However, we have verified that, even in the case of a multi-energy spectrum, the effect of collisions is qualitatively similar to the single-energy scenario. This is consistent with the results reported in the literature for fast flavor conversions~\cite{Shalgar:2020xns}. 
The relaxation of this assumption leads to the possibility of energy changing collisions, which can be incorporated in our formalism. Another possibility that arises in a multi-energy system is the one of a collision term that is energy preserving, but has a strength that depends on energy. The latter can be easily implemented numerically. We have performed such a calculation and find that the results are qualitatively similar to the single energy case.

An effect of practical importance is that the inclusion of the collision term suppresses the formation of structures at small angular scales. This reduces the number of angle bins required to reach convergence in numerical simulations. Contrary to what one may naively expect, the numerical simulations including  collisions are computationally much more efficient than in the case without collisions in our framework. This may play a major role in the ultimate goal of tracking the  flavor evolution in the context of  hydrodynamic simulations of compact objects. However, it should be noted that we solve an initial value problem with a unique mathematical solution. When the full feedback between the collision term and neutrino flavor evolution is included, the problem becomes a time-dependent boundary problem, and there would be no guarantee that it has a unique  solution. 
Our findings also highlight that a self-consistent analysis  cannot be complete without taking in to account in the interplay of three effects simultaneously: neutrino flavor evolution, advection and collisions.

An important question regarding fast flavor conversions is whether the system reaches a steady-state configuration. The answer to this question continues to elude us, but the inclusion of collisions sheds a new light on the matter.

\acknowledgments
We would like to thank Georg Raffelt and Rasmus S. L. Hansen for helpful discussions. We are grateful to the Villum Foundation (Project No.~13164), the Danmarks Frie Forskningsfonds (Project No.~8049-00038B), the Knud H\o jgaard Foundation, and the Deutsche Forschungsgemeinschaft through Sonderforschungbereich
SFB~1258 ``Neutrinos and Dark Matter in Astro- and
Particle Physics'' (NDM).


\bibliography{coll1.bib}

\end{document}